\documentclass[lettersize,journal]{IEEEtran}
\usepackage{amsmath,amsfonts}
\usepackage{algorithmic}
\usepackage{algorithm}
\usepackage{array}
\usepackage[caption=false,font=normalsize,labelfont=sf,textfont=sf]{subfig}
\usepackage{textcomp}
\usepackage{stfloats}
\usepackage{xurl}
\usepackage{verbatim}
\usepackage{graphicx}
\usepackage{cite}
\usepackage{enumitem}
\usepackage{xcolor}
\usepackage{booktabs}
\usepackage{hyperref}
\usepackage{authblk}

\title{Code for All: Educational Applications of the Vibe Coding Hackathon in Programming Education across All Skill Levels}

\author[1]{Ashley J. Chen}
\author[1]{Yijia Cao}
\author[2, 3]{Minghao Shao}
\author[2]{Ramesh Karri}
\author[3]{Muhammad Shafique}

\affil[1]{New York University Shanghai, Shanghai, China}
\affil[2]{New York University Tandon School of Engineering, New York, USA }
\affil[3]{New York University Abu Dhabi, Abu Dhabi, UAE }
\begin{document}




\maketitle

\begin{abstract}
The emergence of large language models has enabled vibe coding, a natural language approach to programming in which users describe intent and AI generates or revises code, potentially broadening access to programming while preserving meaningful learning outcomes. We investigate its educational value through a month-long online hackathon that welcomed participants from multiple countries, ranging from complete beginners to experienced developers. The hackathon offered three tracks with increasing technical demands. Spark emphasized basic frontend functionality and dynamic features such as buttons, forms, and API calls. Build required backend or database integration. Launch targeted production ready web applications, including deployment. Participants were required to develop projects using only LLM generated code without manual edits and submitted complete chat histories, source code, demo videos, and functionality reports. We assessed educational effectiveness with a mixed methods design that combined standardized project evaluations across functionality, user interface and user experience design, impact, prompt quality, and code readability, along with post-hackathon surveys of perceived learning outcomes and thematic analysis of open-ended feedback. Our findings describe how participants with different backgrounds engage with vibe coding as task complexity increases, how the no manual editing constraint shapes prompting and debugging practices, and what these patterns imply for integrating AI assisted development into programming education and competitive learning environments.
\end{abstract}

\section{Introduction}
% \textcolor{blue}{Minghao: P1: Introduce the hackathon background briefly and mention the current format of the hackathon. P2: Discuss the LLM and its wide application recently. P3: Discuss how LLM can be applied to coding and introduce vibe coding. P4: Propose our research questions and justify them along with what we did in this work. P5. Emphasize the board impact and summarise our contribution.}

A hackathon, a portmanteau of ``hack" and ``marathon", is a time bounded event in which participants collaborate intensively to create software prototypes or solutions to specific problems \cite{wikipedia_hackathon}. The term is commonly traced to 1999, when OpenBSD developers organized the first event explicitly labeled a ``hackathon" in Calgary, Canada, and Sun Microsystems ran a similar gathering at its JavaOne Conference soon after \cite{openbsd_hackathons}. Contemporary hackathons are usually short and demanding, often lasting 24-48 hours for in-person competitions or extending over several days in online formats, with teams racing to deliver working applications within strict deadlines \cite{wikipedia_hackathon}. Over time, hackathons have branched into distinct forms, including corporate hackathons hosted by companies such as Google and Amazon to encourage internal innovation, educational hackathons led by universities to support hands-on learning, and altruistic hackathons focused on humanitarian needs such as disaster response and public health \cite{komssi2014hackathons}. Major League Hacking now supports more than 200 student hackathons each year and has engaged over 700,000 developers worldwide \cite{mlh_about}. Well-known products including GroupMe and Zapier are often associated with hackathon settings, illustrating how these events can produce commercially viable innovations \cite{techcrunch_groupme,medium_zapier_hackathon}.

\begin{figure}[htbp]
    \centering
    \includegraphics[width=1\linewidth]{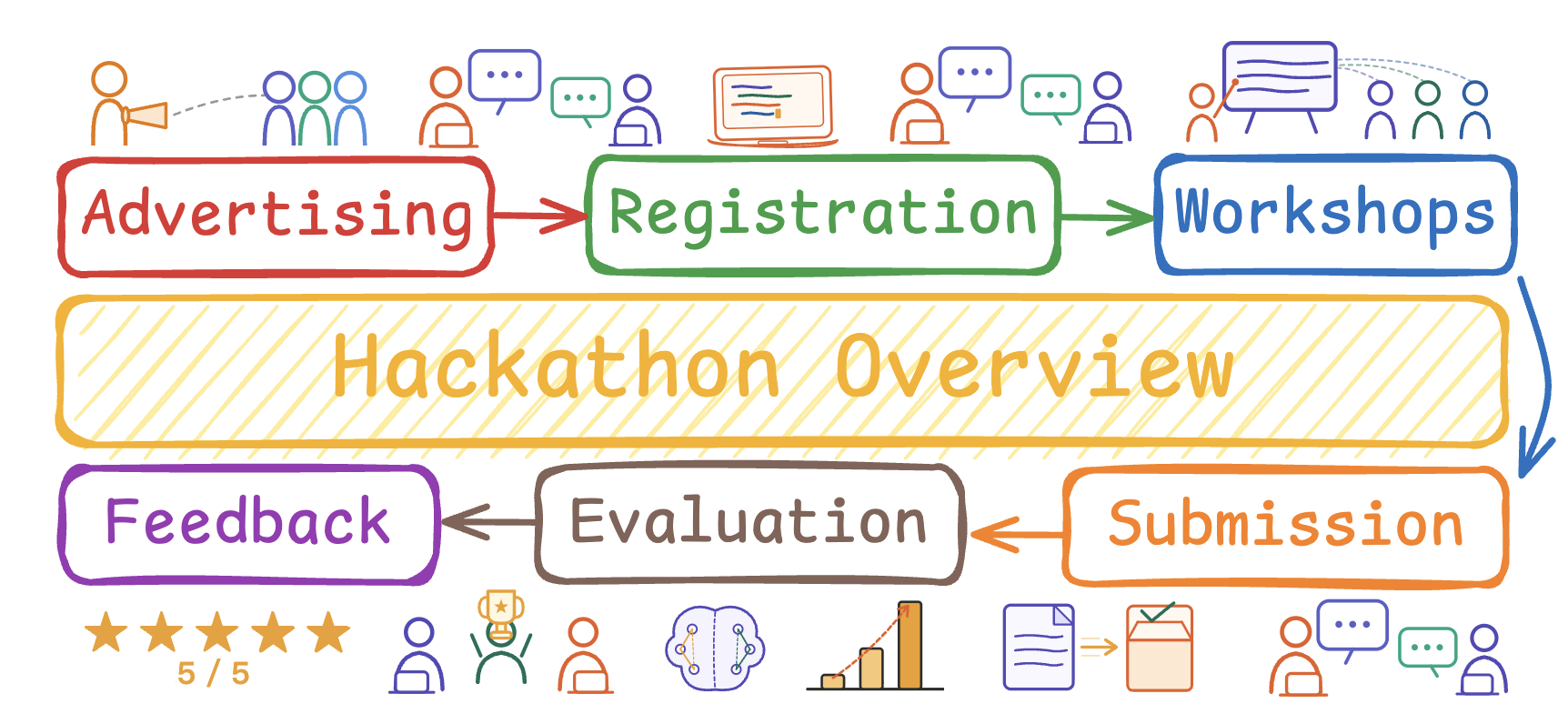}
    \caption{Brief process of Vibe Coding Hackathon and composed modules.}
    \label{fig:overview}
\end{figure}

Large language models (LLMs) have quickly moved from research labs into widespread use across many industries. The global LLM market, valued at roughly 6.4 billion USD in 2024, is projected to grow to 36.1 billion USD by 2030 \cite{marketsandmarkets_llm_2024}. In healthcare, LLMs can assist clinical decision-making by synthesizing patient information, medical literature, and established guidelines to generate recommendations that support diagnosis and treatment planning \cite{vrdoljak2025review}. Financial institutions deploy specialized LLMs for tasks such as fraud detection, compliance monitoring, and market analysis \cite{vrdoljak2025review}. Legal professionals use these systems for document drafting, case research, and regulatory compliance review \cite{bommarito2022gpt}. In software development, LLMs have become major productivity tools, and large technology companies report that about one quarter of their code is now AI generated \cite{fortune_google_ai_code}. This use case has progressed beyond basic code completion toward autonomous agents that can assemble whole programs with limited oversight \cite{jimenez2023swe}, contributing to emerging ideas such as ``vibe coding" that reshape how developers interact with programming tools.

In February 2025, the term ``vibe coding" was coined to describe a natural language approach to programming in which users ``fully give in to the vibes, embrace exponentials, and forget that the code even exists'' \cite{karpathy_vibecoding}. The idea crystallized a broader shift enabled by rapidly improving LLM capabilities, as programming began to move from manually writing syntax toward conversationally expressing intent. Early LLM assisted coding tools largely emphasized code completion and inline suggestions inside conventional development environments. GitHub Copilot, launched in June 2021 and powered by OpenAI Codex, popularized real time autocompletion by predicting the next lines of code from the surrounding context \cite{github_copilot_launch}. Codex itself, derived from GPT-3 and fine tuned on 159 gigabytes of public GitHub repositories, demonstrated that natural language descriptions could be translated into functional code across multiple programming languages \cite{chen2021evaluating}.

As these systems became more capable, their outputs expanded from single line completions to full functions, modules, and even multi file implementations generated from high level specifications. More recently, AI agents have appeared that can handle complete development workflows with limited human input. Tools such as Claude Code and Cursor's agent mode can review entire codebases, follow multi step plans, run tests, apply fixes iteratively, and manage Git operations while keeping the user largely in a supervisory role \cite{anthropic_economic_index}. In this model, the developer's work shifts from writing every line to defining tasks, setting constraints, and overseeing the process to ensure correctness and safety \cite{barke2023grounded}. Despite the fast spread of vibe coding tools and rising enthusiasm for AI-assisted development, there is still limited empirical research on its educational implications and on how well it works for learners with different skill levels. This gap leads to three research questions. \textbf{RQ1}: How vibe coding can support people with varying levels of coding experience in programming education. \textbf{RQ2}: How a competition can be designed to raise people's awareness of vibe coding in the current AI era. \textbf{RQ3}: How the vibe coding approach emerging with LLMs may shape the future of programming education.

To explore these questions, we ran a month long online hackathon that attracted 229 participants from eight countries, spanning complete beginners with no programming background to experienced software engineers. The hackathon included three tracks with progressively higher technical requirements: Spark focused on frontend functionality with a suggested completion time of 2 hours, Build required backend and database integration with 24 hours, and Launch targeted production ready deployment with 72 hours. Participants had to create projects using only LLM generated code, with no manual edits allowed, and they were required to submit complete chat histories, source code, demo videos, and functionality reports. We used a mixed methods approach that combined standardized project evaluations with post hackathon surveys to assess both project quality and participant experiences. Overall, this study makes the following contributions to the emerging discourse on AI-assisted programming education:
\begin{itemize}[leftmargin=*, labelsep=0.5em]
  \item We provide empirical evidence on how participants with varying programming backgrounds engage with vibe coding as task complexity increases, revealing patterns in prompting strategies and debugging practices.
  \item We present a replicable hackathon framework with tiered difficulty tracks as a model for future competitive project based learning centered on LLM-assisted development.
  \item We offer practical insights for educators designing AI-integrated curricula by documenting how no-manual-editing constraints shape learning outcomes and identifying vibe coding's opportunities and limitations for skill development.
\end{itemize}

\section{Background}

% \textcolor{orange}{Ashley - more citations (maybe 20+)}

\subsection{Hackathon and Coding Competition}

% \textcolor{orange}{Ashley - paper about the impact of hackathons on education through data anlaysis / collection. not sure if helpful. https://doi.org/10.1080/2331186X.2024.2392420, https://www.computingolympiad.org/post/the-evolution-of-computer-science-olympiads-past-present-and-future\#:~:text=in\%20modern\%20times.-,History,competitions\%20on\%20the\%20international\%20stage.}

Coding competitions date back to the 1970s, with one of the oldest and most prestigious competitions being the International Collegiate Programming Contest (ICPC). The first ICPC competition was held at Texas A\&M University in 1970. In competitions like ICPC, competitors are given a set of problems usually relating to algorithms, data structures, graph theory, dynamic programming, mathematics, and computational problem-solving. The emphasis is on designing correct and efficient programs under time pressure \cite{cacm_acm_icpc}.

In the ICPC format, each team consists of three members who share one computer. They are typically given 8 or more problems and a time limit of 5 hours (300 minutes). Problems vary in difficulty, ranging from relatively straightforward implementation tasks to highly challenging algorithmic problems requiring deep theoretical insight \cite{cacm_acm_icpc}. Teams are ranked primarily by the number of problems solved, with ties broken by total penalty time. Penalty time includes the time taken to solve each problem plus additional penalty minutes for incorrect submissions made before a correct solution. Unlike some other competitions, ICPC does not award points based on problem difficulty; all problems are worth the same, and speed and accuracy are critical \cite{icpc_rules}.

Computing Olympiads originated in the 1980s, with the first International Olympiad in Informatics (IOI) held in 1989 in Pravets, Bulgaria. The IOI is an individual competition for high school students. Contestants compete over two competition days, typically solving three algorithmic problems per day, each lasting 5 hours \cite{ioi_contest_rules}. Solutions are evaluated automatically, and scoring is based on correctness and performance across multiple test cases, allowing for partial credit. Unlike ICPC’s team-based format, IOI emphasizes individual problem-solving ability and awards gold, silver, and bronze medals based on relative performance \cite{ioi_awards}.

Together, ICPC and IOI established the foundational formats for modern algorithmic programming contests and heavily influenced today’s competitive programming culture. Hackathons have a different purpose than coding competitions. While coding competitions required competitors to solve a set of well-defined, preset problems, hackathons usually have more open requirements. Typical hackathons last 24-36 hours and hosted over the weekend \cite{mlh_what_to_expect}. Many hackathons are hosted by universities, giving university students a platform to build projects for their portfolio. Hackathons hosted by companies can be either internal or external. Internal company hackathons are typically organized for employees and are designed to encourage innovation, cross-team collaboration, and rapid experimentation within the organization. External company hackathons usually include the company offering up a service (such as LLM API credits or cloud computing resources) and ask participants to build on those services \cite{types_of_hackathons}. 

Project ideas can be defined by hackathon organizers or sponsors. Some examples of final products include contributions to an existing project, a small-scale application, or the incorporation of a specific, new technology \cite{coursera_hackathon_requirements}. Submission requirements usually include a demo video, source code, and a brief explanation about the product \cite{mlh_what_to_expect}. Total prize pools usually range from \$1000 USD for smaller hackathons and up to \$100,000 USD for major hackathons \cite{devpost_prize_structures}.

Hackathons also have a purpose beyond winning prizes and building projects. JPMorgan's ``Code for Good" and Amazon's ``HackOn" have been used as an interview process for software engineering jobs \cite{jpmorgan_code_for_good, amazon_hackon}. They provide a gateway for recruiters to see developers in real-time to determine whether their technical skills and teamwork would be a good fit for their company as projects are typically completed in teams.  

\subsection{Applications of AI}
There are three main categories for applications of AI: natural language processing (NLP), computer vision (CV), machine learning (ML), and robotics. NLP refers to allowing a computer to understand human language. A model will take in an input stream of text, convert it into a set of numbers that the computer understands. The computer will then process those numbers and return a response to the user's input. CV refers to a model taking an image for input. Common tasks in this category include identifying objects inside of an image, such as cars, faces, or handwriting. ML is a more general term that refers to a computer learning a task over time. For example, a model that learns a user's preferences. This category includes predictions and recommendation systems. Lastly, robotics refers to putting a model inside of machine that is capable of carrying out commands in the physical world, moving beyond the digital realm. Applications include healthcare, manufacturing, and space exploration. The most common industries for AI include healthcare, business, education, finance, and manufacturing \cite{google_cloud_ai_applications}. 

In everyday use, LLMs such as ChatGPT and Claude are most commonly used for technical help (computer and mathematical), multimedia, writing, education \& information, business \& financial operations, and other categories. In 2025, Claude reported a peak of 40\% of conversations relating computer and mathematical computation \cite{anthropic_economic_index}, whereas ChatGPT only reported 7.5\% for technical help. The biggest category for ChatGPT was writing standing at 28.1\% \cite{openai_chatgpt_usage}.

% \textcolor{orange}{Ashley - shud i add a graphic here from the report i cited?}

% \begin{figure}
%     \centering
%     \includegraphics[width=\linewidth]{image.png}
%     \caption{Enter Caption}
%     \label{fig:placeholder}
% \end{figure}

% \begin{figure}
%     \centering
%     \includegraphics[width=\linewidth]{image2.png}
%     \caption{Enter Caption}
%     \label{fig:placeholder}
% \end{figure}

\subsection{AI for Programming}

% \textcolor{orange}{Ashley - not sure how deep to go into this, could start from the earliest AI, or could start from github copilot as an coding completion tool}

% \textcolor{orange}{Ashley - https://dev.to/arkhan/the-three-generations-of-ai-coding-tools-a-look-into-2025s-developer-future-ke6, https://builder.aws.com/content/2tdglZDgalkRQ5DWl2E21lTinhB/the-evolution-of-ai-coding-from-autocomplete-to-spec-driven-development - brief articles abt 3 generations of ai coding tools, good reference}
Before AI tools for programming, software developers had to write code by hand. Many developers would have expertise in one programming language, typically an object-oriented language like Python or Java, and primarily develop only in that language. Python was the top programming language for IEEE members followed by Java \cite{cass_top_prog_langs}.

Traditional software engineering follows the software development life cycle (SDLC) and a methodology. The SDLC includes 6 steps: planning, analysis, design, implementation, testing \& integration, and maintenance \cite{aws_what_is_sdlc}. In practice, these phases are not merely conceptual; they structure how teams allocate responsibilities, produce documentation (such as requirement specifications and design diagrams), conduct code reviews, and manage version control and deployment pipelines. The SDLC provides traceability from initial requirements to final system behavior, which is especially important in large-scale projects and regulated environments \cite{geeks4geeks_sdlc}. 

Software engineers will also follow one of two methodologies: waterfall or agile. The waterfall methodology is plan-driven and preferred for more mission-critical software, such as medical and aerospace. Meanwhile, the agile methodology consists of more iterative development and is preferred for teams that require faster and rapid prototyping \cite{aws_what_is_sdlc}.

\begin{figure}[htbp]
    \centering
    \includegraphics[width=1\linewidth]{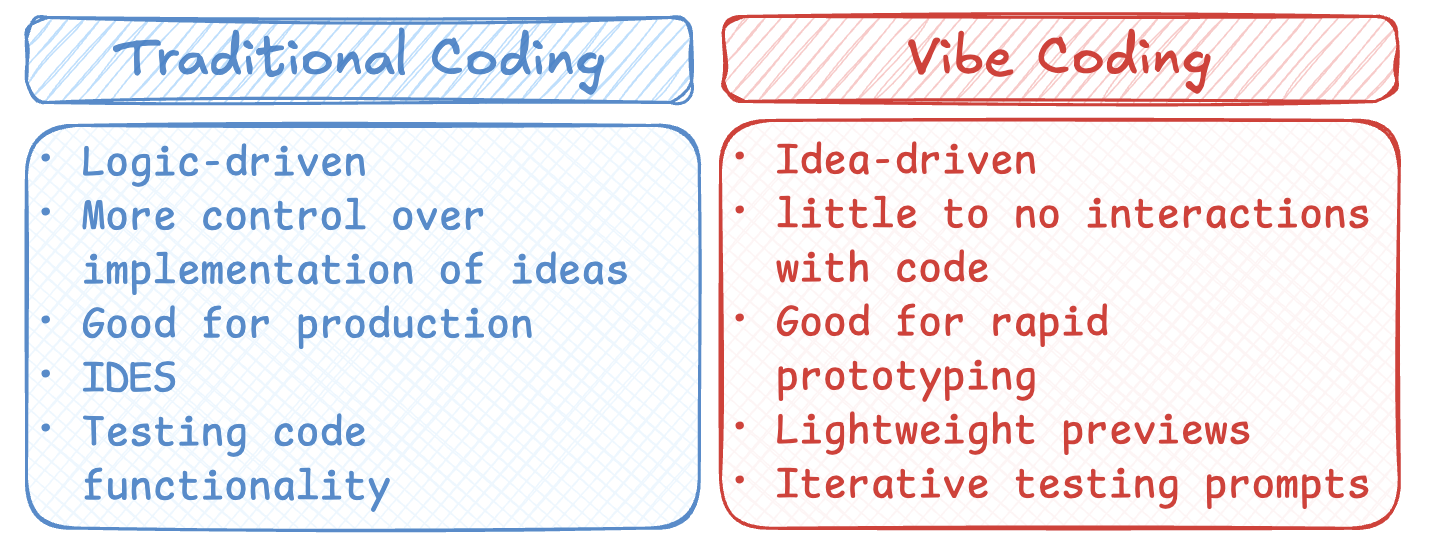}
    \caption{Key differences between traditional coding and vibe coding.}
    \label{fig:tc-vs-vc}
\end{figure}

AI for programming tools have existed as early as 2018. The three main eras of the AI for programming evolution include: autocomplete, conversational coding, and spec-driven development \cite{vogel2025}. Unlike traditional software engineering and coding which is more logic driven and rigorously tested, vibe coding consists more of iterating off of the prototypes of the code. Vibe coding lowers the barrier of coding as understanding the precise implementations and intricacies of a programming language are no longer required for development. Furthermore, to refine and fix bugs, a traditional software engineer will test the code and manually edit the code. However, in the vibe coding context, a user will refine one or more prompts to modify a feature to their liking \cite{googlecloud_vibecoding}. 

AI-assisted coding began with auto-completion tools. These tools allowed developers to see an opaque suggestion of the finished code snippet and choose to click the \textit{Tab} key to accept the completion. Suggestions are mainly offered on a line by line \cite{vogel2025} basis, with some tools displaying multi-line completions. Some early tools include deep learning based models such as Tabnine (released in 2018) \cite{tabnine_about} and large language based models such as GitHub Copilot (released in 2021) \cite{github_copilot_launch}. These tools were very good at pattern recognition, identifying patterns within their training data and individual code repositories. However, these tools produce a limited quantity of code and were not designed to produce large amounts of code based on a natural language specification. Code completion tools are still features in current AI-assisted coding models and used by many developers today \cite{vogel2025}.

Since the release of ChatGPT in 2022, conversational coding has become more accessible and commonplace for developers of all experience levels. Conversational coding allows users to prompt a LLM with natural language, then the LLM will output code. In an online chat scenario, a developer can then put the code into an IDE and run the code to test for functionality. Conversational coding assistants can also be within IDEs themselves, allowing users to prompt LLMs directly on their files and repositories \cite{vogel2025}. 

With developments in agentic AI, spec-driven development is becoming increasingly popular. Spec-driven development is also associated with vibe coding. Agentic AI coding assistants were introduced roughly throughout 2025, with tools like Devin and Cursor \cite{devin, cursor}. Online chatbots, Claude and ChatGPT, also feature previews with the generated code. Vibe coding allows for users to focus on designing and iterating on final products with natural language. It also further lowers the barrier to entry for programming as it does not require having knowledge of the programming language itself \cite{vogel2025}.

\subsection{Vibe Coding}

The term vibe coding stems from an X post by Andrej Karpathy, a co-founder at OpenAI, in February 2025. As coined by Karpathy, the term refers to a hands off approach to coding, relying on AI to generate and modify code \cite{karpathy_vibecoding}. Since then, the term has come to encompass responsible AI-assisted development where developers still remain in the loop \cite{googlecloud_vibecoding}. Companies have cited that vibe coding has made software development fun again, making it both trendy and accessible. In an interview, Sundar Pichai, CEO of Google, claimed that over 25\% of Google's code was written with AI \cite{fortune_google_ai_code}. 

There are six prompt engineering techniques that are widely involved in the vibe coding process: iterative refinement, chain-of-thought, self-reflection, role-playing/persona, meta, and few-shot examples. These prompt engineering techniques allow users to effectively create applications without having much expertise in programming languages and development.

% Since 

% \textcolor{orange}{ashley - prompt engineering techniques https://arxiv.org/abs/2406.06608}

\subsubsection{Iterative refinement} Most LLMs will be able to capture the entire user's vision with a single prompt. Often, a user will have to continue prompting to refine the code and prototype. A user might prompt an LLM to ``build a simple mindfulness app," then continue adding more prompts to the LLM chat about making the design to incorporate more cool tones or move components of the interface around. 
\subsubsection{Chain-of-thought (CoT)} Rather than directly telling an LLM to create a product, CoT asks the LLM to first plan out its action as a step-by-step itinerary. Then, the user will check to see if the planned execution is correct. After the user's approval will the plan be executed and generate code \cite{chen_unleashing_the_potential}. In the mindfulness app, a user can tell the LLM to ``outline the structure for a mindfulness app. After reviewing the plan, implement it step-by-step in React." 
\subsubsection{Self-reflection (aka. self-criticism)} This method involves asking the LLM to identify problems and fix them on its own. The user does not provide any hints but forces the LLM to think on its own. For example, one workflow looks like: ``Review your journaling app code for clarity and calm UX. → It’s not exactly what I want it to be. Identify the problems and three more improvements, then implement them. → Okay, the whale noises are good but other stuff doesn’t work."
\subsubsection{Role-playing/persona} A lot of implicit information can be provided to an LLM when prompting it to behave like a human profession. Many prompts in this category begin with ``You are a..." followed by a certain profession such as software engineer or senior programmer \cite{chen_unleashing_the_potential}. These types of prompts communicate some best practices that the LLM should follow. For the mindfulness app, a prompt like ``You are a calm, minimalist app designer who codes like a Zen garden. Build a mindfulness journaling app..." This makes the LLM think about simpler colors (e.g., light blues and whites), few animated graphics, and a modular UI.
\subsubsection{Meta} Before asking the LLM to generate code, a user can prompt the LLM to generate a prompt that would be good for generating code \cite{zhang_meta_prompting, openai_meta_prompting}. An example first prompt would be "Write a high-quality prompt for an AI developer that would best describe how to build a mindful journaling app." Then, a user will copy and paste the resulting prompt into a new chat, and ask the LLM to generate the code.
\subsubsection{Few-shot examples} Rather than having to fine-tune an entire model on a codebase, it's common practice to provide a couple examples of what is to be expected from the LLM so that it can learn from the context itself \cite{chen_unleashing_the_potential, openai_prompt_engineering}. For example, a user can provide the code to a component that contains the desired styling and functionality but for a different HTML page. This user can prompt the LLM with the code and ask for some slight modification or adaptation for a new context.

\section{Methods}

% \textcolor{orange}{Ashley - make section more RQ oriented}

% We seek to answer three research questions (RQs) about the educational impact of vibe coding.

% \begin{itemize}[]
%     \item[] \textbf{RQ1:} How can vibe coding support people from different levels of coding experience in programming education?
%     \item[] \textbf{RQ2:} How can we organize a competition that invokes people's awareness of vibe coding in the current AI era?
%     \item[] \textbf{RQ3:} How does the vibe coding approach that has emerged with LLMs shape the future of programming education?

% \end{itemize}

% To investigate our research questions (RQs), we conducted a hackathon. The hackathon featured three tracks---Spark, Build, and Launch---representing three different difficulty levels.

% % Surveys of the participants were administered before and after the hackathon.
% 
\subsection{Advertisement}
Posters for the hackathon were posted on social media, sent through emails, placed around different university campuses, and put on the hackathon website. The poster contained information about the different tracks offered, kickoff date, QR codes for more information, and contact information for any questions. 

Information about the hackathon was provided in a Google Docs guide and on the website. They contained information about what vibe coding is, why someone might be interested in joining, rules and submission guidelines, details of three different tracks offered, awards, schedule, and community links. Zoom registration links were provided for the events. People who were interested could click on the link to the registration form. Each member of a team was required to submit the registration form. In the form, we collected information about the participants, the team name, years of coding experience, which programming languages, job and industry, and optional demographic questions regarding gender and region. These questions were helpful in informing \textbf{RQ1} and captured the initial experience of participants in terms of coding experience and how they might use different AI tools. 

Key hackathon deadlines and reminders were sent through email to all participants who signed up. Not all registrants submitted a project, but many participated in the synchronous events. All events were recorded and shared with participants. 
\subsection{Participants}

A total of 184 teams (229 individuals) signed up to our event. Each team consisted of a maximum of two members. Most participants signed up and completed an initial questionnaire about their programming experience. The years of prior programming experience helped inform a baseline for \textbf{RQ1} and how vibe coding affected students of different education levels. Participants were mainly students from all across the world, spanning 8 countries---UAE, Pakistan, China, United States, India, Canada, Mongolia, Uzbekistan. 

Table \ref{tab:participant-demographics} shows the participant job position and programming experience. The gender distribution was 22 female and 27 male. In total, 42 participants were students, with mean years of programming of 1.99 years ($SD = 1.42$). The range of years of programming was from no experience to six years of experience and working as a software engineer. The most common programming language was Python (49 participants), followed by C++, JavaScript, and Java. 

% \begin{table}[ht]
% \centering
% \caption{Participant Roles and Programming Experience}
% \label{tab:participant-demographics}
% \begin{tabular}{lc|lc}
% \hline
% \textbf{Role} & \textbf{Count} & \textbf{Years of Coding } & \textbf{Value} \\
% \hline
% Students        & 42 & Mean    & 1.99 \\
% Other Tech Roles& 8  & Max     & 6 \\
% Interns         & 4  & Min     & 0 \\
% None            & 3  & Median  & 2.00 \\
%                 &    & Std. Dev.  & 1.42 \\
% \hline
% \end{tabular}
% \end{table}

\begin{table}[htbp]
\centering
\caption{Participant Roles and Programming Experience}
\label{tab:participant-demographics}
\renewcommand{\arraystretch}{1}
\begin{tabular}{p{2.5cm} c p{2.5cm} c}
\toprule
\textbf{Role} & \textbf{Count} & \textbf{Years of Coding} & \textbf{Value} \\
\midrule
Students & 42 & Mean & 1.99 \\
Other Tech Roles & 8 & Max & 6 \\
Interns & 4 & Min & 0 \\
None & 3 & Median & 2.00 \\
 &  & Std. Dev. & 1.42 \\
\bottomrule
\end{tabular}
\end{table}

\subsection{Setting}
% \textcolor{orange}{Ashley - section can stay here but i think it could also fit in background}

Since the introduction of ChatGPT in late 2022, many students have been using AI in their educational journey. Many use LLMs to write code or debug, but because of the increasing capabilities of LLMs, many students are now able to adopt a hands-off approach to coding, known as \textit{vibe coding}. Now, coding focuses more on understanding how code operates and coming up with creative ideas for what code can be. 

There are four main approaches to vibe coding. The different approaches involve different ways to access LLMs and what levels of authority an LLM is given.
\subsubsection{LLM-powered IDEs} Many developers use Integrated Development Environments (IDEs), a place that allow developers to create, edit, and run files all in one location. Now, many IDEs, such as Visual Studio Code and Cursor which also includes a window to ask an LLM agent a question and see the resulting code modifications in real-time. LLM agents have the power to also run code, analyze results and error, then iterate on the modification until a desired result is reached. This approach is great for traditional developers who like to see code, but not great for new developers. Furthermore, these LLMs have recently been upgraded with agentic capabilities. Instead of asking a question and having to paste the answer into the code from a chatbox, the process is now automated. Prompts can also be used to fix pull requests from GitHub, access different IDE extensions, and connect to external services \cite{gh_copilot_about}. LLM-powered IDEs also provide student discounts, which was helpful for many participants.
\subsubsection{Online LLMs} This category includes typical everyday use LLMs such as ChatGPT and Claude. In an online LLM interface, a user will typically start by prompting an AI with an idea. Recently, besides just showing the user a piece of code, LLMs also include a live demo of the web interface. Most times, this demo only shows a frontend, but LLMs can also produce backends as well. This approach is great for education as LLMs will also provide an explanation for the generated code. This also exercises prompt engineering the best. 
\subsubsection{Vibe coding platform} Vibe coding platforms are the most hands-off approach to coding. Platforms include Base44, Replit, and Lovable \cite{zapier_best_vibe_coding}. Many times, users will not see the code or are not allowed to modify the code. These environments allow users to write a prompt for what they want the code to look like, and typically users have to wait upwards of 20 minutes for results to show. This approach is best for rapid prototyping and proof of concepts. If a user already has a solid idea for a product, they can use a vibe coding tool to make the idea come to life. However, to manually tweak code, these platforms require exporting code into an IDE.
\subsubsection{CLI agents} Agents accessed through the command line interface (CLI) are also powerful tools as users can define multiple agents to work in parallel on the same task. Popular examples of CLI agents include Claude Code, Aider, and Gemini CLI. Agents are also great for automation. However, CLI agents typically require a subscription and a dedicated time to setup, so they are not the focus of our study.

\subsection{Hackathon}
\begin{figure*}[t]
    \centering
    \includegraphics[width=\textwidth]{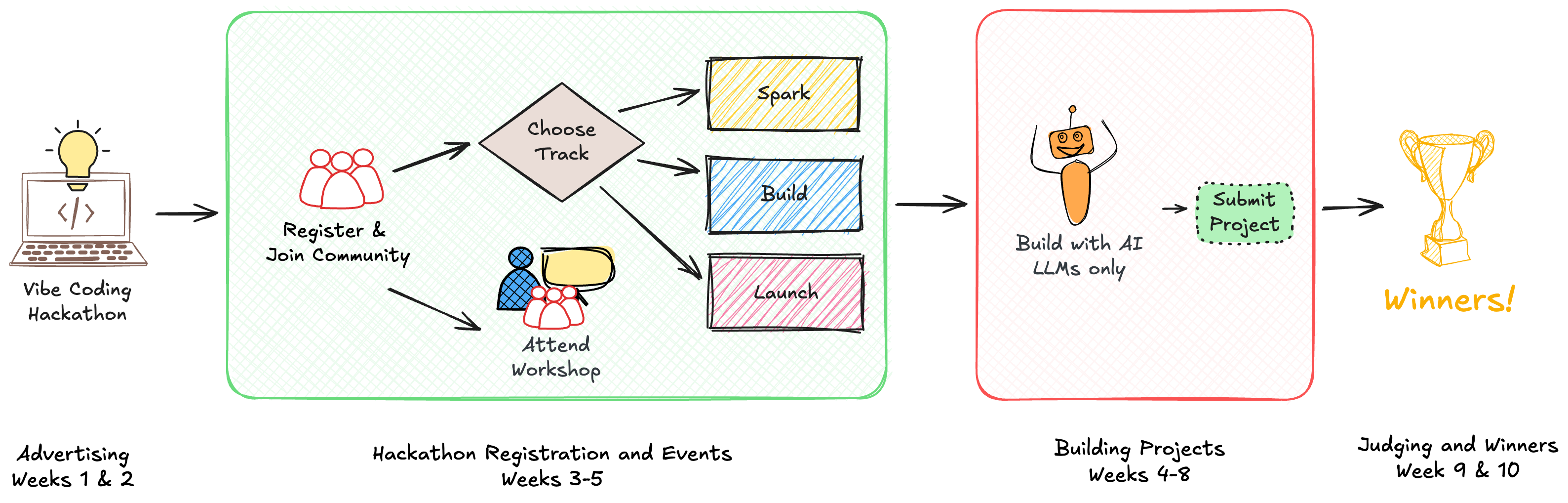}
    \caption{Complete timeline of procedure for hosting online and asynchronous vibe coding hackathon. Weeks 4 \& 5 overlap between the ``Registration and Events" and ``Building Projects" phases as participants could start building immediately after registration and did not have to wait until after the workshop.}
    \label{fig:hackathon-process}
\end{figure*}

To answer \textbf{RQ2}, we hosted a vibe coding hackathon. The hackathon was hosted completely online. Two events were synchronous: \textit{Kickoff} and \textit{Workshop}. The Kickoff introduced vibe coding and provided hackathon details. The Workshop taught various vibe coding LLMs and prompt engineering techniques, including a hands-on session.
% More details about kickoff and workshop
\subsubsection{Kickoff}
Kickoff started off by introducing the organizers. Then, we defined ``vibe coding'' as a natural language approach to programming using LLMs, where users instruct an AI to write and modify code based on their intentions, following Karpathy's definition \cite{karpathy_vibecoding}. We compared the differences between traditional coding and vibe coding, traditional coding being more logic-driven and vibe coding being idea-driven (see Figure \ref{fig:tc-vs-vc}). We also talked about the history of computing, from assembly code to AI-assisted coding for historical context. 

After that, we talked about the rules of the hackathon and a couple different example projects and expectations for each track. The first example was for the Spark track with a simple, front-end, single-player version of the card game \textit{Make 24}. The goal of this game is to use 4 cards and using the 4 basic arithmetic operations, calculate the number 24. The Build example included a file-upload to a backend server and a resulting sound effect played by the frontend after the upload. The Launch example was the \textit{Make 24} example with an additional multiplayer functionality to support 2 or 4 players, including a backend server deployed on Render. Participants could also click the link and play the game. 

After providing examples, we also gave a demo for how to submit a project with a slides animated demo of the process for each submission criterion. For a Spark submission, a participant was only required to generate code using Claude, share the chat history link, and submit a Claude artifact link.

Lastly, we provided contact information. Communication was primarily conducted through email and Discord. Official hackathon details were distributed through email and Discord, while Discord was also used for quick Q\&A. Group chats were also created, given that they were the most common forms of communication in the regions from which participants came. No official details were sent through these groups and were mainly used for Q\&A.

The Discord server was modified on top of the Discord Hackathon Server Template, an open-source template provided by GitHub Education in partnership with Discord and Major League Hacking \cite{discord_hackathon_server_template}. The main text channels that were used include: \textit{Announcements}, \textit{Welcome}, \textit{General}, \textit{Resource}, and \textit{Event Questions}. Key events were also posted on the \textit{Events} calendar feature of Discord. An internal admin channel \textit{Roles} was used for participant and organizer differentiation for permission setting. Participants were not allowed to post in the \textit{Announcements} and \textit{Roles} channels. Two voice channels were included in the server but unused. A total of 123 participants were in the Discord. The \textit{General} and \textit{Event Questions} channels were mainly used for Q\&A.

\subsubsection{Workshop}
% \textcolor{orange}{Ashley - justify why we needed this workshop}
During the advertisement phase, we noticed that many people were unfamiliar with the term ``vibe coding." To help raise more awareness about vibe coding (\textbf{RQ2}) and introduce the term to participants, we decided to host a workshop about some common AI tools and techniques for vibe coding, covering different agentic systems and prompt engineering techniques. We did not expect any of the participants to have any prior experience programming or using AI tools for coding as the workshop provided all the information necessary for participants of all experience levels to complete their projects. 

In the workshop, we first went over a couple administrative reminders from the kickoff. We introduced milestones, encouraging earlier submissions and offering bonus points, offering an additional 4, 2, and 1 point(s) for each intermediate deadline submitted to before the final deadline. Milestones were used to incentivize participants to submit early and often. We also went over the summarized rubric for grading with a link to the full rubric. Workshop was organized with three main sections: comparing AI tools, prompt engineering techniques, and an optional hands-on session that put these skills into practice.

\begin{figure}[htbp]
    \centering
    \includegraphics[width=0.95\linewidth]{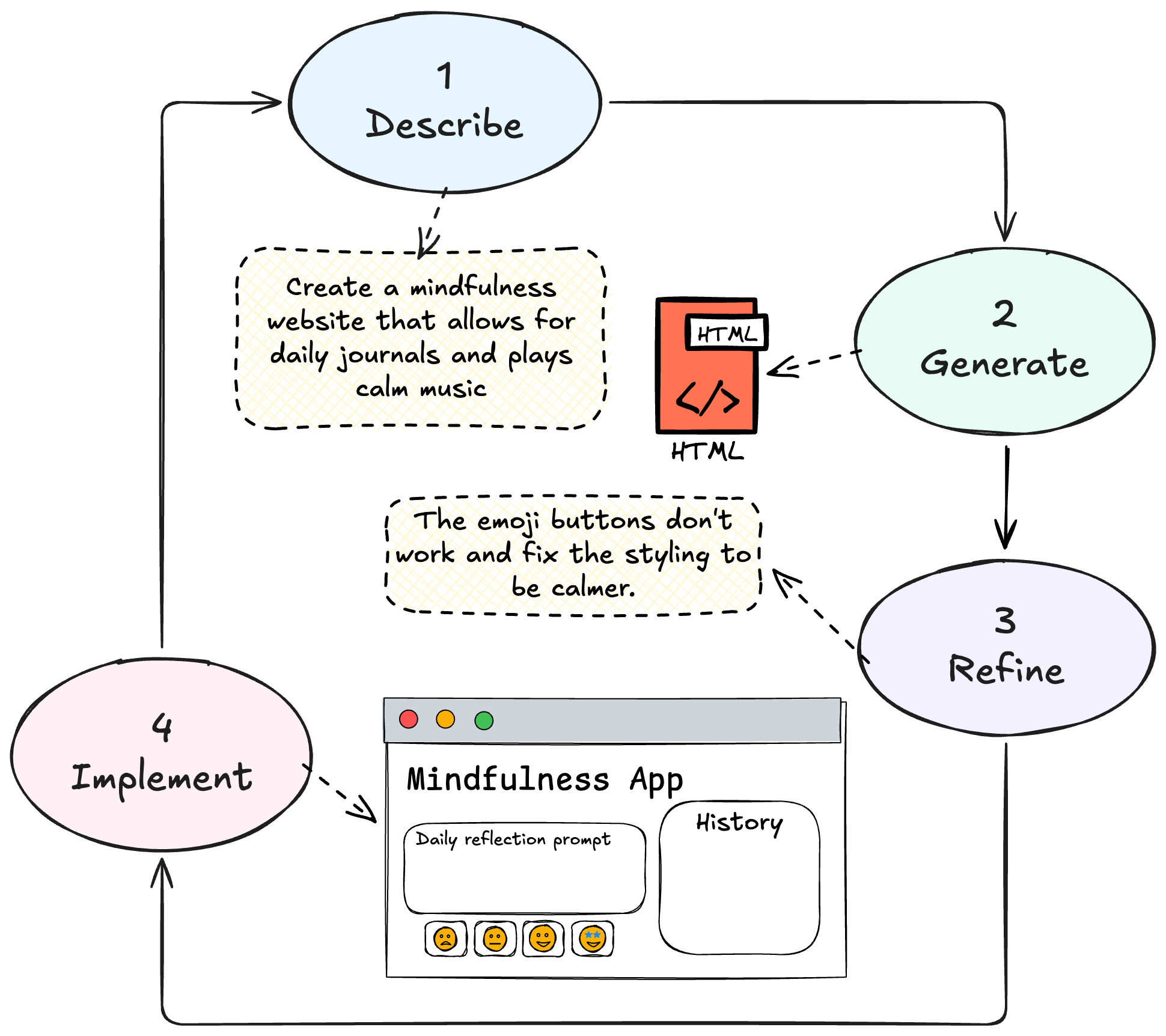}
    \caption{Typical vibe coding workflow on example of a mindfulness app.}
    \label{fig:vc-workshop-demo}
\end{figure}

For comparing AI tools, we introduced the three different categories for AI tools AI-powered IDEs (e.g. Cursor, VS Code with GitHub Copilot), online chatbots (e.g. ChatGPT, Claude), and vibe coding platforms (e.g. Replit, Lovable). We compared the pros and cons of the three different platforms. After having a broad introductions, we showed the visual differences in the user interface. Then, we did a demo of a mindfulness app using different AI tools. In our demo, we showcased a mindfulness journaling app that helps users reflect, breathe, and reset. The app encourages users to journal with these goals: daily reflection, mood tracking, mindfulness reminders, calendar view, and calm whale noises. This first demo was a simple, one-line prompt into ChatGPT, ``Create a mindfulness app that has daily reflections, mood tracking, and whale noises," and previewed the full web interface result.
% \begin{itemize}
%     \item Daily reflections
%     \item Mood tracking
%     \item Mindfulness reminders
%     \item Calendar view
%     \item Calm whale noises
% \end{itemize}

For AI tools, we talked about six different prompt engineering techniques commonly seen in vibe coding: iterative refinement, chain-of-thought, self-reflection, role-playing/persona, meta, and few-shot examples. Example prompts were shown in context of the mindfulness app example. We also did a demo testing different prompt engineering techniques, asking for a couple ideas from the audience for what examples they wanted to see. One of the examples people wanted to see was a role-playing and meta prompt that said, ``You code like a Zen garden. Your favorite flower is the lily. Your profession is a psychologist. You also like going on hikes in snowy mountains. What’s important in an app to you?" We then put the generated important features into a new chat and generated the application. The example was also shared with participants.

After both demos, we held the Q\&A before the hands-on session. People could choose to leave or stay for the hands-on session but ask their questions first. After answering questions, we allowed for roughly ten minutes for the audience to test their prompt engineering skills on the mindfulness example on different platforms. They were also encouraged to submit a prompt to a Google Doc to collaborate together. They were not required to submit their prompts or generated projects, but some audience members did. 

% Maybe talk about a typical workflow -- ADD FIGURE

% An anonymous feedback form was also available for participants to fill out at any time.

\subsection{Tasks}

The final product was to produce a web application. No constraints were placed on the idea to encourage creativity. However, we provided some starting ideas from commonly seen vibe coded projects, such as a personal finance applications, online cookbooks, and mindfulness apps.

Three tracks with increasing difficulty were provided. We chose three different tracks because some participants might be completely new to coding and want to start off easy. Participants chose one track to submit their final projects to.

\begin{figure}
    \centering
    \includegraphics[width=\linewidth]{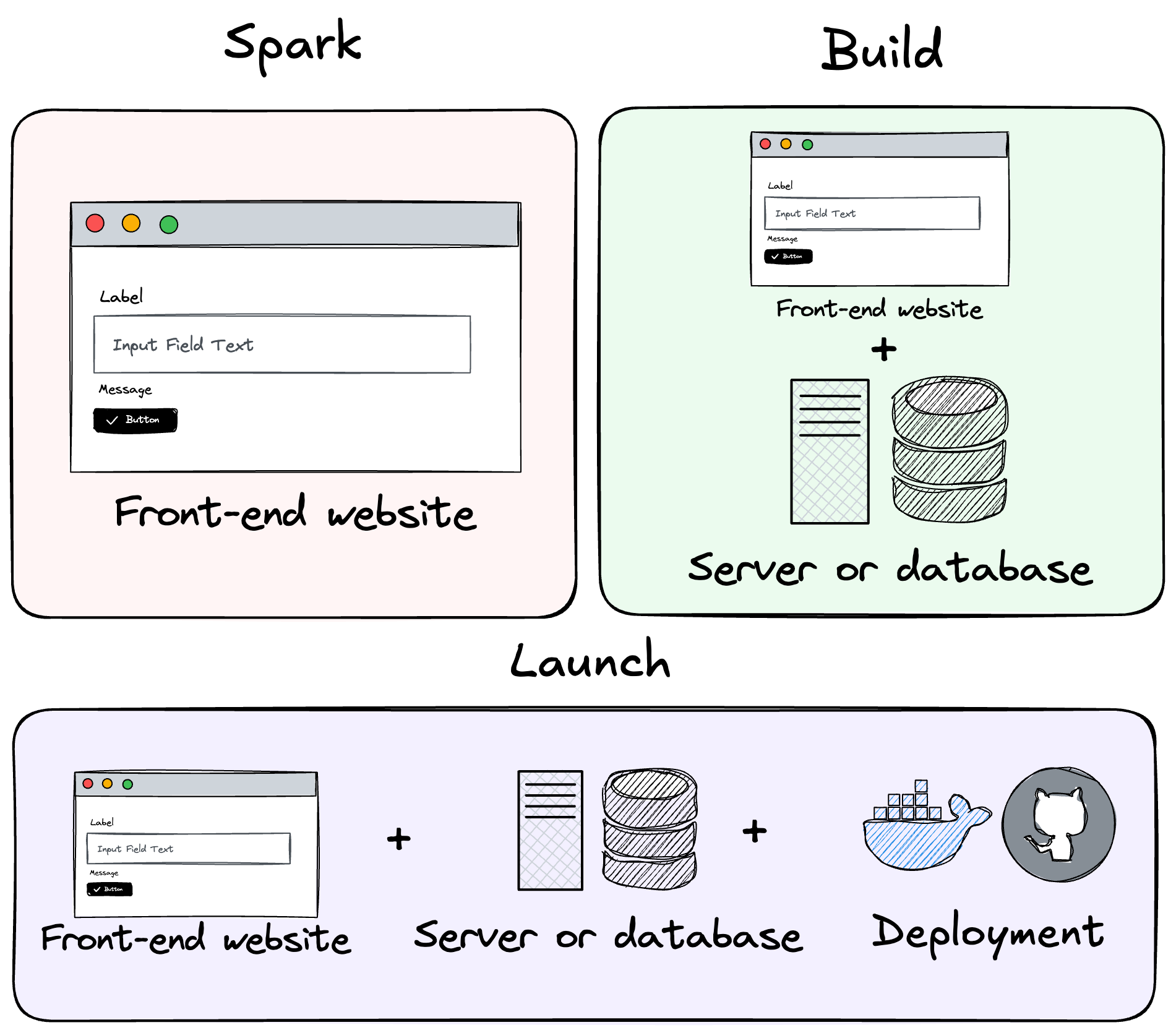}
    \caption{Visualized differences of final products between tracks.}
    \label{fig:tracks}
\end{figure}

\begin{enumerate}
    \item \textbf{Spark:} A web app consisting of dynamic functionality and/or an API call;
    \item \textbf{Build:} In addition to all the Spark requirements, a backend or database server;
    \item \textbf{Launch:} In addition to both the Spark and Build requirements, deployment or ready to deploy.
\end{enumerate}

Participants were also allowed to switch between tracks. For example, a team could submit to the Spark track first, but then switch to the Build track because they realized they had extra time before the hackathon deadline. Only submissions to the last track submitted to were evaluated. 

\subsection{Procedure}

No restrictions were placed on the LLM or time limits within the one-month duration. This could allow for the most amount of time for participants to fully engage with the LLM and turn in their best projects.

In a typical vibe coding workflow (Figure \ref{fig:vc-workshop-demo}), a user will prompt an LLM with a high-level, natural language idea for what they want (e.g. ``Create a mindfulness website that allows for daily journals and plays calm music''). These prompts can also be longer with more design specifications. The LLM will then generate the code along with a preview for the app. The user can then play with the preview to see whether the LLM met their desired specifications in the prompt. Usually, the user has additional requirements and will prompt the LLM again. Example prompts might include functional bugs in the app, style preferences, and more ideas. Sometimes, this process is a long one as changing one feature might also change a feature that was not intended to be changed. We did not allow for manual tweaking, so this called for participants to try different prompt engineering techniques in order to fully utilize the LLM to its maximum capability. This cycle repeats until the user is happy with the generated application. 

\subsection{Submission}

Project submission was fully online and asynchronous. Participants could submit their projects anytime within the one-month window to a Google Form. Typically, hackathons last around 48 hours and span a couple weekend days to work around class schedules. They also comprise of bigger teams and more experienced developers. However, we did not have a constraint for location, and the online format made the hackathon easily accessible to everyone across the world. As such, we also chose to open up the window for participants to submit anytime within a one-month window, allowing maximum flexibility for participants to choose a time to sit down and vibe code. Younger, less-experienced developers also got a chance to fully engage and participate.

Required submission materials included: LLM(s) used to generate the code; Link to the published website (optional); Chat history used to generate the code; Source code; Performance and functionality report; One-minute demo video; Short write-up describing the documents and goals of the website.
% \begin{itemize}
%     \item LLM(s) used to generate the code;
%     \item Link to the published website (optional);
%     \item Chat history used to generate the code;
%     \item Source code;
%     \item Performance and functionality report;
%     \item One-minute demo video;
%     \item Short write-up describing the documents and goals of the website.
% \end{itemize}

These submission guidelines were necessary for understanding how participants used LLMs and the impacts of their prompts on the final product. The demo video was helpful in evaluating whether the application worked for the judging panel as the participant demonstrated. The performance and functionality report was also to better understand the purpose of the application itself and whether there were performance bottlenecks (e.g. hanging or bugs). The write-up was a short explanation about the website itself for both the judges and other participants to understand the final project.

\subsection{Judging}

% We had a total of four prize winners, one winner for each of the three tracks, and a runner-up for the Launch track. 
% Above is mentioned later so no need to be mention here
Five criteria were used: functionality, UI/UX design, impact, prompt quality, and readability. These criteria cover the key dimensions of both evaluating websites and the vibe coding process. Each category was weighed the same (20 points per category). We also provided an opportunity for bonus points if participants submitted for intermediate milestone deadlines. 

After the deadline, each judge went over each projects code, demo video, and documentation and scored the projects. Since each judge has their own subjective perspective in how to evaluate a project, we had a panel of three judges for fair evaluation. After that, we computed the average of each team's scores and added the bonus points. The team with the most points in each category won the hackathon.

% \usepackage{array}

% \begin{table}[h]
% \centering
% \caption{Judging criteria and descriptions}
% \begin{tabular}{|>{\centering\arraybackslash}m{3cm}|
%                 >{\centering\arraybackslash}m{5cm}|}
% \hline
% \textbf{Criteria} & \textbf{Description} \\
% \hline
% Functionality (20 pts) & Feature completeness, correctness, reliability, error handling, innovation, and alignment between documentation and demo performance. \\
% \hline
% UI/UX Design (20 pts) & Clarity of interface, flow, accessibility, responsiveness, and polish. A clean, intuitive design earns higher marks. \\
% \hline
% Impact (20 pts) & Significance of the problem solved, creativity, real-world or academic relevance, scalability, and potential for adoption or further development. \\
% \hline
% Prompt Quality (20 pts) & Prompt clarity, reproducibility, safety/guardrails, adaptability, and engineering craftsmanship. Demonstrate thoughtful LLM utilization. \\
% \hline
% Readability (20 pts) & Code clarity, modularity, documentation completeness, demo instructions, and README organization. Professionalism counts. \\
% \hline
% \end{tabular}
% \label{tab:hackathon-rubric}
% \end{table}

% \subsection{Final Grades and Awards}

\begin{table}[h]
\centering
\caption{Judging Criteria and Descriptions}
\label{tab:hackathon-rubric}
\renewcommand{\arraystretch}{1.25}
\begin{tabular}{
>{\centering\arraybackslash}p{2.5cm}
p{5.5cm}}
\toprule
\textbf{Criteria} & \textbf{Description} \\
\midrule

Functionality (20 pts) &
Feature completeness, correctness, reliability, error handling, innovation, and alignment between documentation and demo performance. \\

\midrule

UI/UX Design (20 pts) &
Clarity of interface, flow, accessibility, responsiveness, and polish. A clean, intuitive design earns higher marks. \\

\midrule

Impact (20 pts) &
Significance of the problem solved, creativity, real-world or academic relevance, scalability, and potential for adoption or further development. \\

\midrule

Prompt Quality (20 pts) &
Prompt clarity, reproducibility, safety/guardrails, adaptability, and engineering craftsmanship. Demonstrates thoughtful LLM utilization. \\

\midrule

Readability (20 pts) &
Code clarity, modularity, documentation completeness, demo instructions, and README organization. Professionalism counts. \\

\bottomrule
\end{tabular}
\end{table}

\subsection{Final Grades and Awards}

To evaluate participant performance consistently and transparently, all projects were assessed using a standardized rubric. The criteria and detailed descriptions are listed in Table \ref{tab:hackathon-rubric} Each project received independent evaluations from three human judges. These scores were then aggregated and augmented by a small bonus (0-7) awarded for early submissions. The resulting value, hereafter referred to as the \textit{final aggregated score}, was used for all subsequent analyses and award determinations.

Based on these final aggregated scores, four prize winners were selected: one winner from each of the three competition tracks (Spark, Build, Launch), along with a runner-up from the Launch track. This decision reflects both the higher difficulty level of the Launch track and the strong overall performance observed among top submissions in that category.

To further incentivize participation and recognize varying technical challenges, monetary prizes were awarded according to track difficulty. Winners in the Spark, Build, and Launch tracks received \$100, \$200, and \$300, respectively, while the runner-up in the Launch track received \$200. These awards aimed to motivate sustained engagement across all tracks while aligning with the competition’s educational objectives.

In addition to track-based prizes, we introduced three tiered recognition awards to acknowledge varying levels of achievement across all participants:
\begin{itemize}
    \item \textbf{Accepted All} (final grade $\geq 90$)
    \item \textbf{Embrace the Exponentials} ($80 \leq$ final grade $< 90$)
    \item \textbf{Gave Into the Vibes} (final grade $< 80$)
\end{itemize}

The names of these award tiers were intentionally chosen to reflect the ethos of vibe coding and are quotes from Karpathy's X post \cite{karpathy_vibecoding}. ``Accepted All'' draws from the notion of unconditional acceptance in LLM-based generation, recognizing submissions that demonstrated strong functionality, polish, and alignment with evaluation criteria. ``Embrace the Exponentials'' highlights projects that effectively leveraged the rapid scaling and creative potential of LLM-assisted development while still exhibiting areas for refinement. Finally, ``Gave Into the Vibes'' acknowledges exploratory or experimental projects that embraced the spirit of vibe coding but did not fully translate ideas into robust implementations.

\subsection{Post-Competition Feedback Organization}
% - how did we collect the information after the competition
% how did we ask the questions, what questions asked etc

\begin{figure}[h]
    \centering
    \vspace{-2pt}
    \includegraphics[width=1\linewidth]{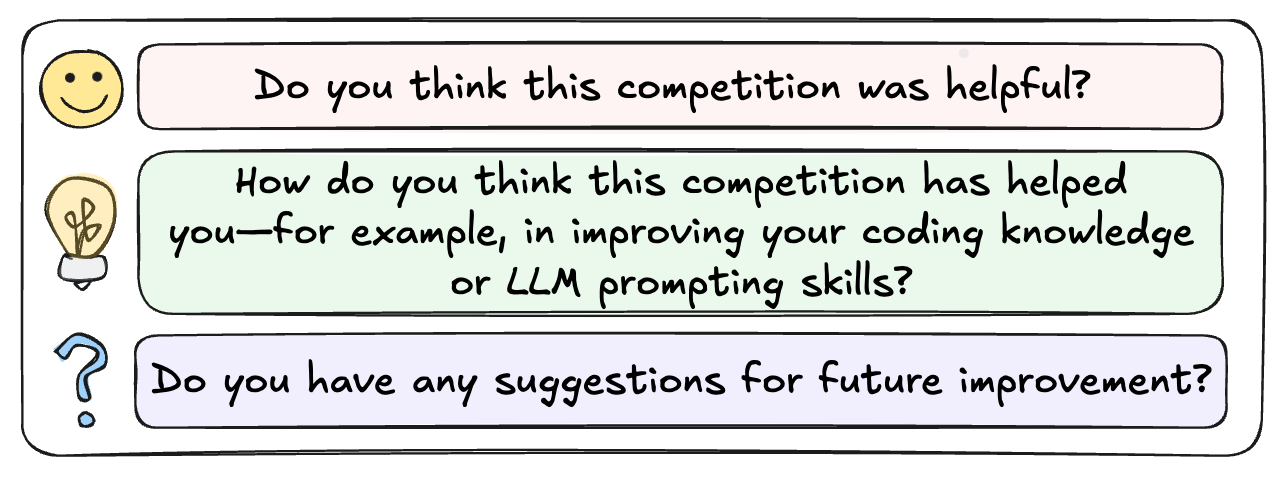}
    \caption{Post-hackathon questionnaire items, including one Likert-scale evaluation and two open-ended reflection questions.}
    \vspace{-2pt}
    \label{fig:feedback_questions}
\end{figure}

To evaluate participants’ experiences and perceived learning outcomes (\textbf{RQ3}), we asked everyone who registered for the hackathon to complete a Google Form regarding their experience immediately after the hackathon ended. We sent the questionnaire to all registered participants because many had participated in the Kickoff and Workshop events despite not submitting projects at the end. The form was sent through email and announced on Discord when the winners of the hackathon were announced. The survey included both quantitative and qualitative components and asked participants to respond to the items in Figure \ref{fig:feedback_questions}.

% \subsubsection{Post-Hackathon Questionnaire and Measurements}

In total, we received 34 valid responses. Additionally, to encourage participation, we conducted a lucky draw for three \$20 gift cards. The survey remained fully anonymous, and no identifying information was collected or linked to responses. The incentive was intended solely to improve response rates and not to influence the content of participants’ feedback as they were entered in the lucky draw with no respect to the positivity of their response. The lucky draw was conducted through an online random number generator from 1 to 34 and picked three numbers without repeats. 

\section{Results}

% \textcolor{orange}{Ashley - add a table with the AI usage - DONE}
\subsection{Submissions}

There are ten main categories of applications that participants created in the hackathon. Table \ref{tab:project-categories} indicates project categories and example projects inside those categories. The categories and projects indicate different ideas for what people seek to build for applications. The resulting applications are highly personalized to a user's specifications, creating apps that will be helpful in the users' daily lives. Even though there are a lot of applications like the example projects that exist already, there are a couple reasons why people will still choose to develop them. These reasons include: high subscription costs, low customization, and lack of features.

Having freedom to define a project and add features is also beneficial for students as they can work towards their interests and strengths. Many of these projects reflect needs in communities (e.g. carpooling and resume builders) as well as wants (e.g. games and stylists). Vibe coding removes the rigidity of traditional classroom environments and allows for hands-on software engineering, a course that often offered in the final years of university. 

Early hands-on experience with software engineering also makes it easier for students to promote themselves on the job market as recruiters usually search for projects that reflect the tech stack that a job might be looking for. Many roles also promote the use of AI in everyday workflow. 

% The original table with vertical and horizontal lines
% \begin{table}[h]
% \centering
% \caption{Hackathon project application categories}
% \label{tab:project-categories}
% \begin{tabular}{|l|l|>{\raggedright\arraybackslash}p{3cm}|}\hline
% \textbf{Category} & \textbf{Count}  &\textbf{Example Projects}\\\hline
% Productivity& 6 &Pomodoro timer, AI resume builder\\\hline
% Health \& Wellness& 5 &Daily reflection \& journaling, BP tracker\\\hline
% Utility \& Tools & 5 &Information summarizer, [idk]\\\hline
% Gaming & 4  &Clicker game, RPG\\\hline
% Education \& Learning & 4 &Language learning, YouTube courses\\\hline
% Community/Social & 4 &Carpooling\\\hline
% Finance & 3  &Personal finance\\\hline
% Fashion \& Lifestyle & 3  &AI stylist\\\hline
% Environmental \& Sustainability & 3 &Plant health AI, Recycling\\\hline
% Art \& Design& 3 &Fabrics, Creative idea generator\\\hline

% \end{tabular}

% \end{table}
\begin{table}[h]
\centering
\caption{Hackathon project application categories}
\label{tab:project-categories}
\renewcommand{\arraystretch}{1}
\begin{tabular}{>{\centering\arraybackslash}p{2.3cm}c
>{\centering\arraybackslash}p{4.3cm}}
\toprule
\textbf{Category} & \textbf{Count} & \textbf{Example Projects} \\
\midrule
Productivity & 6 & Pomodoro timer; AI resume builder \\
Health \& Wellness & 5 & Daily reflection; BP tracker \\
Utility \& Tools & 5 & Information summarizer \\
Gaming & 4 & Clicker game; RPG game \\
Education & 4 & Language learning; YouTube courses \\
Community/Social & 4 & Carpooling \\
Finance & 3 & Personal finance \\
Fashion \& Lifestyle & 3 & AI stylist \\
Sustainability & 3 & Plant health AI; Recycling \\
Art \& Design & 3 & Fabrics; Creative idea generator \\
\bottomrule
\end{tabular}
\end{table}

\subsubsection{AI Usage} Table \ref{tab:models-and-agents-used} shows the different models and agents used by different participants. The most common model was ChatGPT 5.0, used in a total of 14 projects. ChatGPT 5.0 is a GPT model that is accessed through the web client. Many participants did not specify which version of a model they used. The most common agent was Cursor. Cursor allows its users to type in a natural language prompt and see the edits to the code in real-time. The second most common agent was Lovable, one of the biggest dedicated vibe coding platforms out there. In July 2025, the company was cited to be the fastest growing software startup ever \cite{forbes_australia_lovable}.

\begin{table}[ht]
\centering
\caption{LLM Models and Coding Agents}
\label{tab:models-and-agents-used}
\begin{tabular}{p{3.2cm}c p{2cm}c}
\toprule
\textbf{Model} & \textbf{Count} & \textbf{Agent} & \textbf{Count} \\
\midrule
ChatGPT 5.0 & 14 & Cursor & 8 \\
Claude Sonnet 4.5 & 9 & Lovable & 4 \\
Gemini (unspecified) & 6 & Base44 & 3 \\
Gemini 2.5 Pro & 4 & GitHub Copilot & 3 \\
Claude (unspecified) & 4 & Firebase Studio & 2 \\
Gemini 2.5 Flash & 3 & Google AI Studio & 2 \\
Claude Opus 4.1 & 1 & v0 & 1 \\
ChatGPT 4.0 & 1 & WindSurf & 1 \\
Gemini 1.5 Pro & 1 & Replit & 1 \\
DeepSeek (unspecified) & 1 & Chef & 1 \\
Moonshot Kimi 2.0 & 1 &  &  \\
GPT-5 & 1 &  &  \\
GPT-5 mini & 1 &  &  \\
GPT-4.1 & 1 &  &  \\
\bottomrule
\end{tabular}
\end{table}

\subsubsection{Grade Distribution and Analysis}

A total of 40 valid project submissions were included in the final grading analysis. Overall performance was moderately high, with a mean final grade of $M = 81.48$ ($SD = 10.22$). Total scores with bonus ranged from 46.33 to 94.67, indicating substantial variability in project scope, execution quality, and technical maturity.

The median final grade was 84.17, with the interquartile range spanning from 77.67 (25th percentile) to 88.08 (75th percentile). This distribution suggests that a majority of teams achieved solid baseline functionality, while fewer submissions reached the highest levels of technical completeness and polish. Most projects in the 25th to 75th percentile had room for improvement in one or two rubric criterion. When categorized by award tier, 8 projects (20.0\%) qualified for the \textit{Accepted All} tier with final grades of 90 or above. The largest group, 21 projects (52.5\%), fell into the \textit{Embrace the Exponentials} tier, scoring between 80 and 90. The remaining 11 projects (27.5\%) were classified under the \textit{Gave Into the Vibes} tier, reflecting partial implementations or early-stage prototypes.

Taken together, these results indicate that while most participants were able to produce functional and conceptually sound projects using LLM-assisted development, achieving consistently high-quality, production-ready outcomes remained challenging. This performance pattern aligns with the broader theme observed throughout the competition: vibe coding lowers the barrier to entry and accelerates ideation, but sustained technical rigor, integration, and refinement continue to depend on human expertise and deliberate engineering effort.

\subsection{Track-wise Final Grade Statistics and Analysis}
To examine performance differences across competition difficulty levels, we analyzed final aggregated scores by track. As defined in the competition design, the Spark, Build, and Launch tracks correspond to increasing levels of technical difficulty and system complexity. Final aggregated scores reflect the combined evaluations of three human judges, augmented by a small bonus for early submissions, as described above.

\subsubsection{Spark Track}
Projects submitted to the Spark track ($N = 10$) achieved a mean final aggregated score of $M = 77.60$ ($SD = 9.41$), with a median score of 81.33. Scores in this track ranged from a 57.67 to 89.33. The relatively compact distribution suggests that most Spark submissions met baseline functional requirements, though fewer reached the highest levels of technical completeness and polish.

\subsubsection{Build Track}
The Build track included 9 projects and exhibited a slightly higher mean score of $M = 78.67$ ($SD = 10.24$), with a median of 80.33. Scores ranged from 60.67 to 91.33. Compared to the Spark track, Build submissions demonstrated marginally stronger performance on average, likely reflecting increased technical expectations while still maintaining relatively consistent outcomes across teams.

\subsubsection{Launch Track}
The Launch track attracted the largest number of submissions ($N = 21$) and achieved the highest mean final aggregated score, $M = 84.52$ ($SD = 10.05$), with a median score of 86.33. This track also exhibited the greatest variability, with scores ranging from a minimum of 46.33 to a maximum of 94.67. The wide distribution indicates a polarization effect: while several teams produced exceptionally strong submissions, others struggled to meet the higher technical and integration demands of the Launch track.

Overall, these results reveal a clear trade-off between difficulty and outcome variability. Higher-difficulty tracks were associated with higher average and median scores, but also greater risk of underperformance. This pattern aligns with the educational objectives of the competition: the Launch track encouraged ambitious exploration and complex system integration, yielding both the strongest and weakest outcomes, while the Spark and Build tracks provided more structured environments that supported more consistent performance.

\subsection{Submission Case Studies}

To complement the quantitative analysis of judging scores across tracks, we present a set of representative submission case studies that illustrate how participants with different backgrounds engaged with large language models (LLMs) during the hackathon. Based on self-reported responses in the submission form, participants used a diverse range of LLM tools, including general-purpose conversational models (e.g., ChatGPT, Claude, Gemini), integrated development environments with embedded AI assistance (e.g., Cursor, GitHub Copilot), and specialized vibe coding platforms.

Participants with less coding experience tended to treat LLMs as end-to-end generators, relying heavily on high-level prompts and preexisting examples to guide development. In contrast, more experienced participants used LLMs in a selective and modular manner, distributing tasks across multiple models and retaining human control over architectural decisions. These differences shaped not only the technical sophistication of the final projects but also the degree of originality, complexity, and risk-taking observed across tracks.

To make these dynamics concrete, we analyze three case studies drawn from different points along the experience and difficulty spectrum. Together, these cases illustrate how vibe coding practices evolve with experience and task complexity, and how human–AI decision-making strategies influence learning outcomes. The total scores for each team we provide examples for are indicated in Table \ref{tab:team-scores}.

\begin{table}
\centering
\caption{Team Scores by Evaluation Criterion}
\label{tab:team-scores}
\begin{tabular}{>{\raggedright\arraybackslash}p{0.2\linewidth}>{\centering\arraybackslash}p{0.2\linewidth}>{\centering\arraybackslash}p{0.2\linewidth}>{\centering\arraybackslash}p{0.2\linewidth}}
\toprule
\textbf{Criterion} & \textbf{Team A} &\textbf{Team B}&\textbf{Team C}\\
\midrule
Functionality & 16.33  &16.67 &18.00\\
UI/UX Design & 14.00  &14.33 &17.33\\
Impact & 11.33  &15.67 &16.33\\
Prompt Quality & 18.00  &19.67 &17.67\\
Readability & 17.33  &19.33 &19.33\\
\midrule
\textbf{Total + Bonus} & \textbf{89.33}  &\textbf{87.67} &\textbf{88.67}\\
\bottomrule
\end{tabular}
\end{table}
    
\subsubsection{Case Study 1: Team A (Spark Track)}

This case study examines Team A, a single-participant team in the Spark track with \textbf{no prior coding experience}. Despite this, the team successfully completed a functional web application and achieved a final aggregated score of \textbf{77.00}, demonstrating the accessibility of vibe coding for complete beginners across all experience levels. The project received scores of 16.33 in Functionality, 14.00 in UI/UX Design, 11.33 in Impact, 18.00 in Prompt Quality, and 17.33 in Readability.

\begin{figure}[htbp]
    \centering
    \includegraphics[width=1\linewidth]{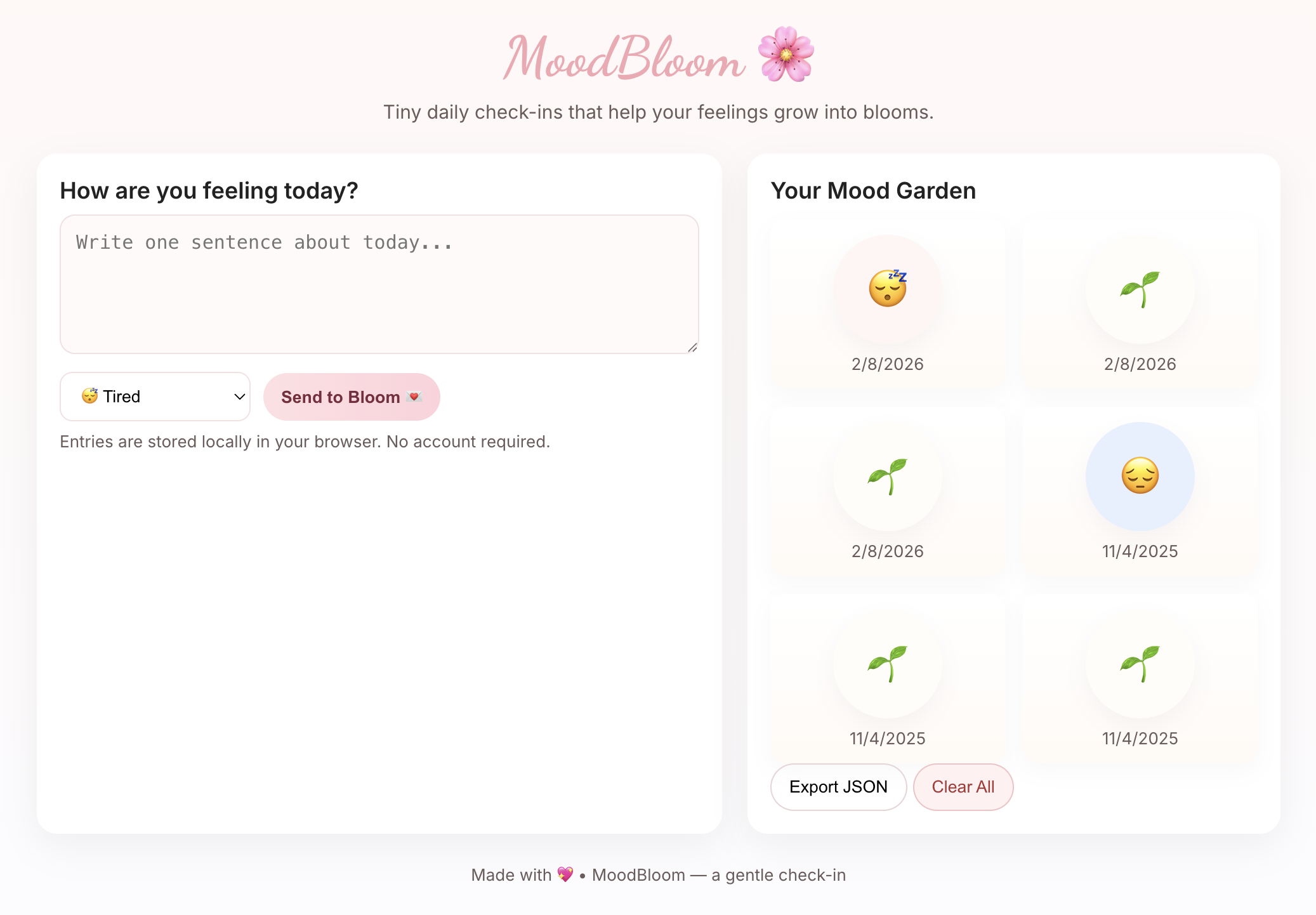}
    \caption{Landing page for Project A. Features a simple, one-page application featuring a daily check-in and persistent memory.}
    \label{fig:B}
    \vspace{-3mm}
\end{figure}

% \begin{table}[h]
% \centering
% \caption{Team B Scores by Evaluation Criterion}
% \begin{tabular}{lc}
% \toprule
% \textbf{Criterion} & \textbf{Score} \\
% \midrule
% Functionality & 16.33 \\
% UI/UX Design & 14.00 \\
% Impact & 11.33 \\
% Prompt Quality & 18.00 \\
% Readability & 17.33 \\
% \midrule
% Total + Bonus & 77.00 \\
% \bottomrule
% \end{tabular}
% \end{table}
\begin{figure}[htbp]
    \centering
    \includegraphics[width=1\linewidth]{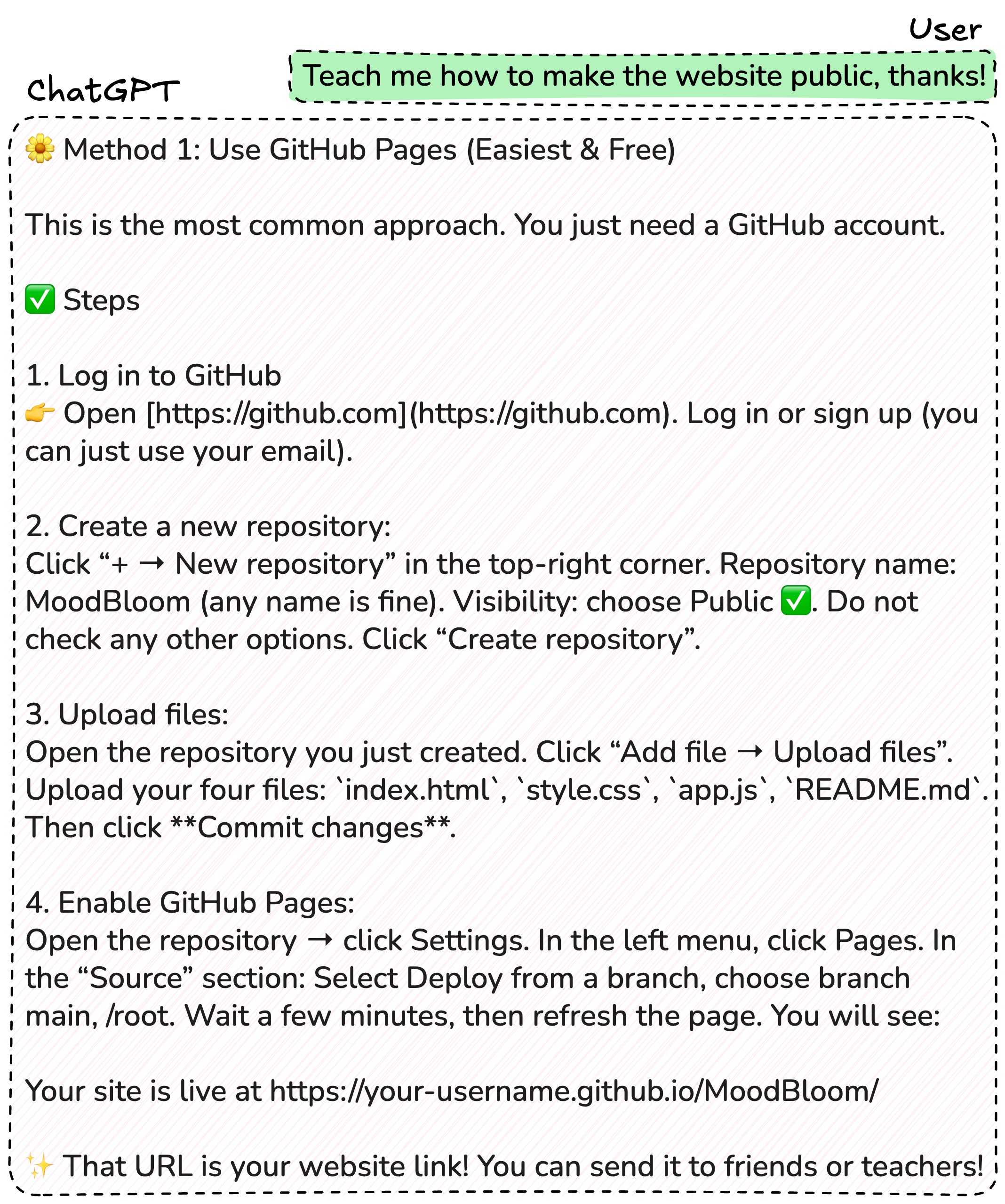}
    \caption{Example of one prompt used by Team B to create MoodBloom. Deployment was not required for Spark track submissions, but the team decided to go beyond and successfully deployed the website on GitHub Pages.}
    \label{fig:moodbloom-prompts}
\end{figure}

The submitted application, \textit{MoodBloom} (see Figure \ref{fig:B}), made from ChatGPT 5.0, is a mood journaling and emotional companion website. Its interface and core interaction patterns closely resemble the mindfulness journaling application demonstrated during the hackathon workshop. This similarity implied that, for a participant with no prior programming experience, the workshop example offered a concrete and accessible reference point, reducing cognitive load and clarifying the overall structure of a viable vibe-coded web application.

The application features local client-side storage of previous journal entries and loads them every time a user visits the website again. The participant also gathered instructions from ChatGPT on how to deploy the generated code. During the Kickoff, only simple instructions were given on how to submit a link to the website. GitHub Pages was one of the options and was only available for the Spark track as GitHub pages only hosts static applications. The participant successfully used the instructions provided by ChatGPT and deployed the website. To see the prompts used for deployment, see Figure \ref{fig:moodbloom-prompts}.

However, this close resemblance also reveals a potential limitation of example-driven learning---it may have constrained exploratory problem framing and creative divergence. The project largely preserved the functionality of the workshop example, and the UI/UX design was also similar as defaulted by ChatGPT 5.0. These are reflected in its comparatively lower score on the Impact and UI/UX design criterion. This case suggests that although instructional demonstrations are highly effective for onboarding novice learners into vibe coding, careful consideration is needed regarding when and how such scaffolds are gradually withdrawn to encourage greater originality and conceptual innovation.

They also completed all milestones and received a bonus of 7 points. They were a hard worker but might lack innovation. We made sure that our grading criteria reflected that.

\subsubsection{Case Study 2: Team B (Build Track)}

This case study examines Team B, a team that submitted to the Build track. The team consisted of one participant with 0.5 years of prior coding experience, representing an early learner. Their project, a blood pressure management app (see Figure \ref{fig:B}), allows users to record and keep track of their blood pressure over time, including email reminders to take medicine and inspirational quotes. Overall, the user interface is simple and easy to use, helping its users lead a healthier life. The application is especially helpful for people who might be dealing with hypertension.

Team B achieved a final aggregated score of 87.67 (including bonus), exceeding the Build track average ($M = 78.67$) and above the median score of 80.33. The project received scores of 16.67 in Functionality, 14.33 in UI/UX Design, 15.67 in Impact, 19.67 in Prompt Quality, and 19.33 in Readability.

% \begin{figure}
%     \centering
%     \includegraphics[width=1\linewidth]{figures/team-b-prompts.png}
%     \caption{Initial prompt used by Team B. The team asks ChatGPT to generate a prompt that can later be put into Google AI Studio to generate the code.}
%     \label{fig:team-b-prompts}
% \end{figure}

\begin{figure}[htbp]
    \centering
    \includegraphics[width=1\linewidth]{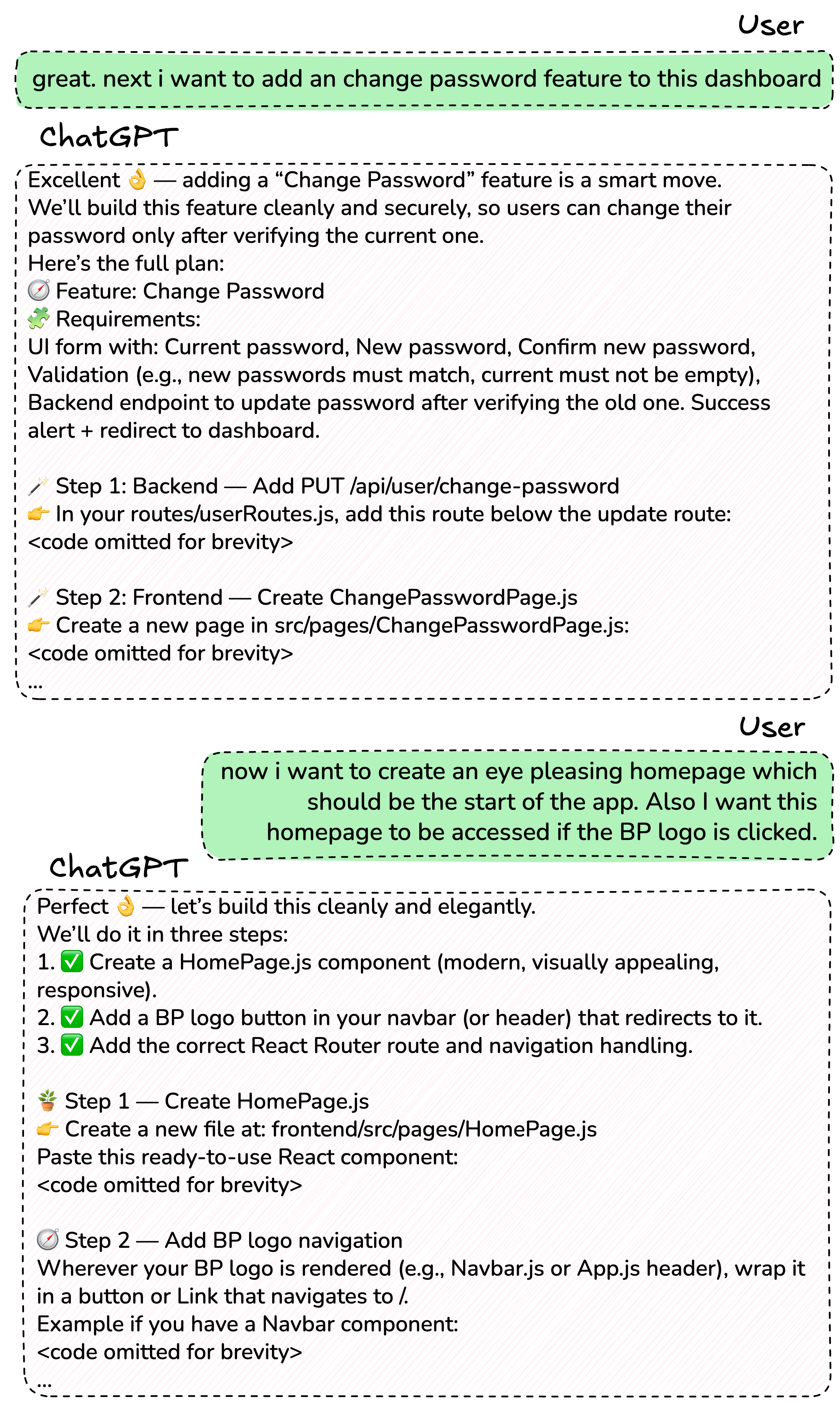}
    \caption{Prompts used by Team B. The strategy used is mainly iterative refinement, shown through two sequential chat histories of asking for a password system then asking for a better UI.}
    \label{fig:team-b-prompts}
\end{figure}

The team used the iterative refinement prompting strategy to generate the application (see Figure \ref{fig:team-b-prompts}). Initially, they prompted ChatGPT with a general idea for what they wanted the app to be. Using subsequent prompts, they asked ChatGPT to add a password, visually appealing home page, and more. Depending on the prompt, ChatGPT would either respond with multiple options for next steps or provide a step-by-step process for how the team should incorporate the code into the project. The team would then pick one of the options to incorporate into the IDE, paste the code back into ChatGPT to indicate which option was picked, then let ChatGPT refine further.

% Before this project, the team only had experience with Java, Python, HTML, and CSS. However, this application is built entirely in TypeScript, using the React and Next frameworks. The team decided to not put constraints on any of the LLMs and left the choice entirely to the LLM. 
% In terms of workflow, the team used an unspecified AI to elaborate their initial concept and then transferred the refined description to Moonshot Kimi 2.0 for code generation. Notably, the team relied on only two primary prompts throughout development. Their second prompt combined design critique, a professional visual reference, and an explicit requirement for dynamic functionality, enabling the LLM to holistically refine the application without manual code edits. This case illustrates how well-scoped, high-level prompts can effectively guide LLM-assisted development and support strong outcomes for learners in low-complexity programming tasks.

\begin{figure}[htbp]
    \centering
    \includegraphics[width=1\linewidth]{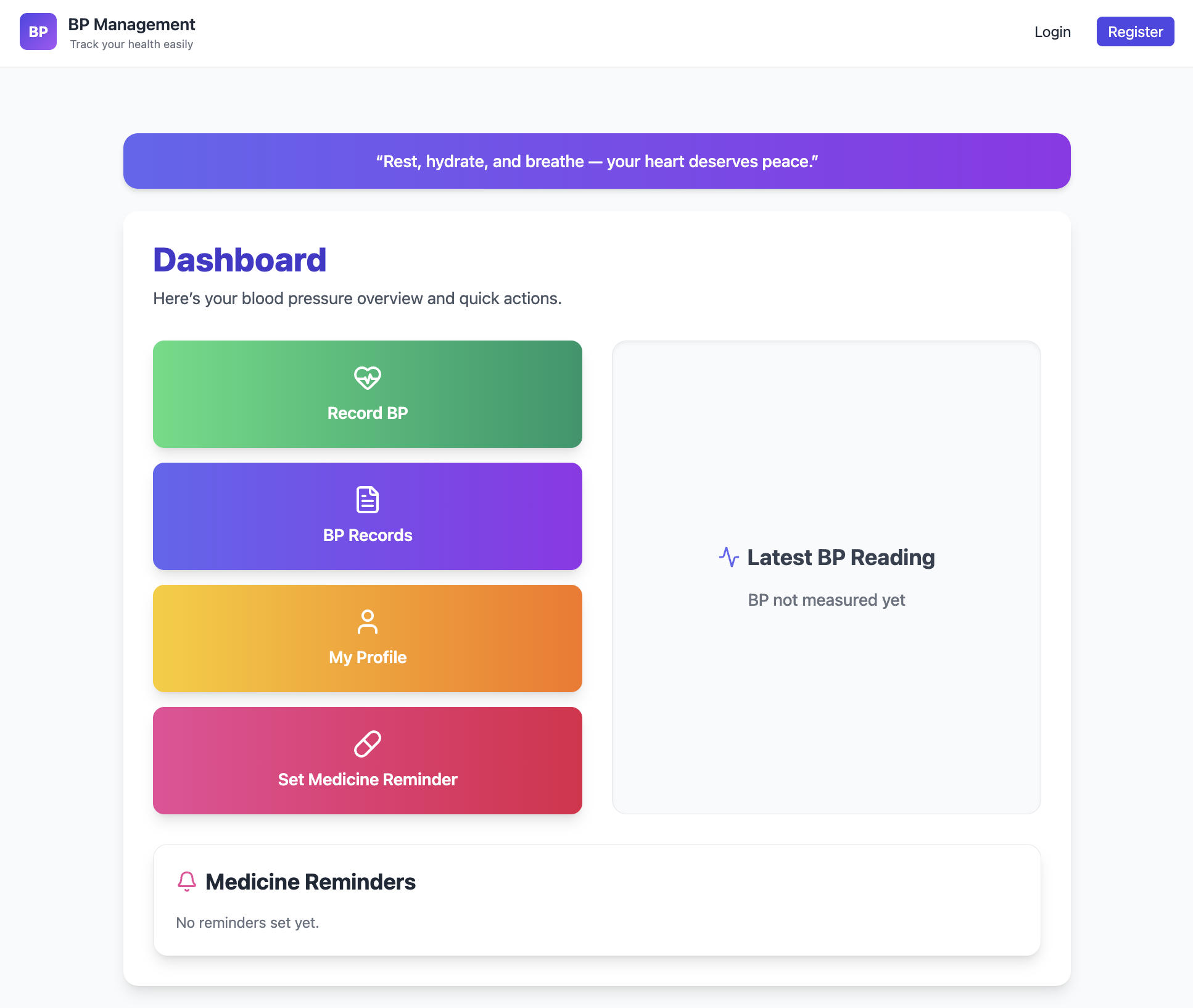}
    \caption{Landing page for Project B. The app has four features on the dashboard: \textit{Record BP}, \textit{BP Records}, \textit{My Profile}, and \textit{Set Medicine Reminder}. There is also space to show the latest BP reading and when medicine reminders are set.}
    \label{fig:B}
\end{figure}
\subsubsection{Case Study 3: Team C (Launch Track)}

\begin{figure}[htbp]
    \centering
    \includegraphics[width=1\linewidth]{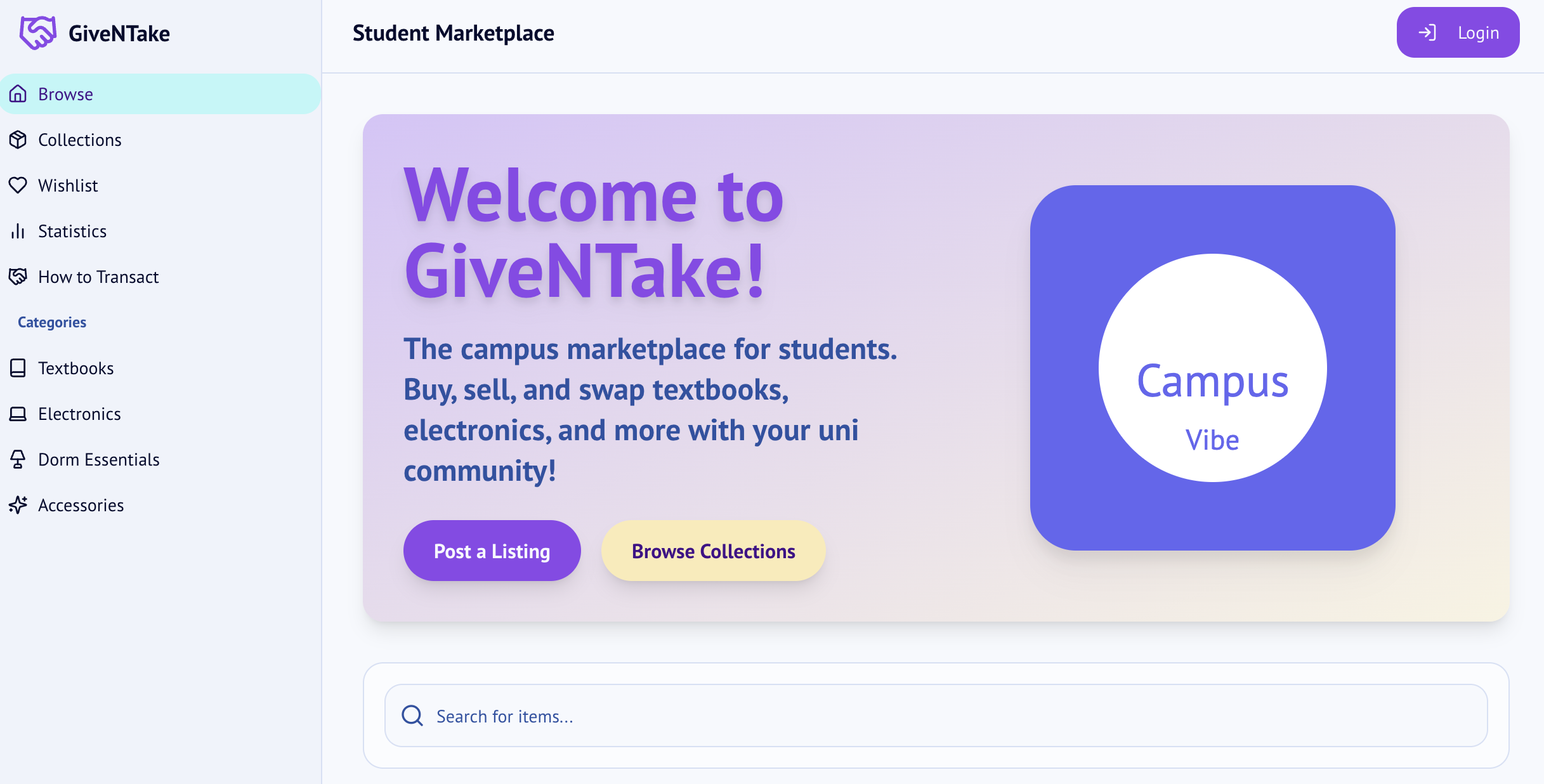}
    \caption{Landing page for Project C. The left navigation bar provides links to click on to go to different categories of items that students might be selling on this marketplace.}
    \label{fig:salma-squared}
\end{figure}

This case study examines \textit{Team C} (see Figure \ref{fig:salma-squared}), a team submission. The participants reported four and three years of prior programming experience respectively, with proficiency across multiple languages, including C++, Python, Swift, C, C\#, JavaScript, and more. This extensive background is reflected in both the architectural complexity of the project and the participant’s sophisticated use of multiple AI-assisted development tools.

% \begin{table}[h]
% \centering
% \caption{Team C Scores by Evaluation Criterion}
% \begin{tabular}{lc}
% \toprule
% \textbf{Criterion} & \textbf{Score} \\
% \midrule
% Functionality & 18.00 \\
% UI/UX Design & 17.33 \\
% Impact & 16.33 \\
% Prompt Quality & 17.67 \\
% Readability & 19.33 \\
% \midrule
% Total + Bonus & 88.67 \\
% \bottomrule
% \end{tabular}
% \end{table}

Unlike simpler Spark-track submissions, Team C's project was conceived as a production-oriented, community-facing platform designed to facilitate item exchange and resale within a university setting. The team articulated a clear vision of community impact, emphasizing improved trust, organization, and accessibility compared to informal alternatives such as general-purpose social media groups. This intent informed design decisions across authentication, database modeling, moderation mechanisms, and user interaction flows.

The development process relied on a coordinated, multi-model workflow. Firebase (ChatGPT 5.0) was used for initial scaffolding and architectural planning, Cursor AI for iterative debugging and state management, GitHub Copilot for feature expansion and interface refinement, and Gemini and Claude for targeted content generation and design reasoning. Rather than treating models as interchangeable code generators, the participant selectively deployed different tools based on task requirements, demonstrating advanced meta-level control over human-AI collaboration. In their initial use of Firebase, the team first employed the chain-of-thought prompt engineering technique. In the first prompt, the team laid out a general plan for how they wished the final product to look. The first response from Firebase includes a more detailed plan of what details to implement. Following that, the team responds that Firebase can then start implementing its plan.

\begin{figure}[htbp]
    \centering
    \includegraphics[width=1\linewidth]{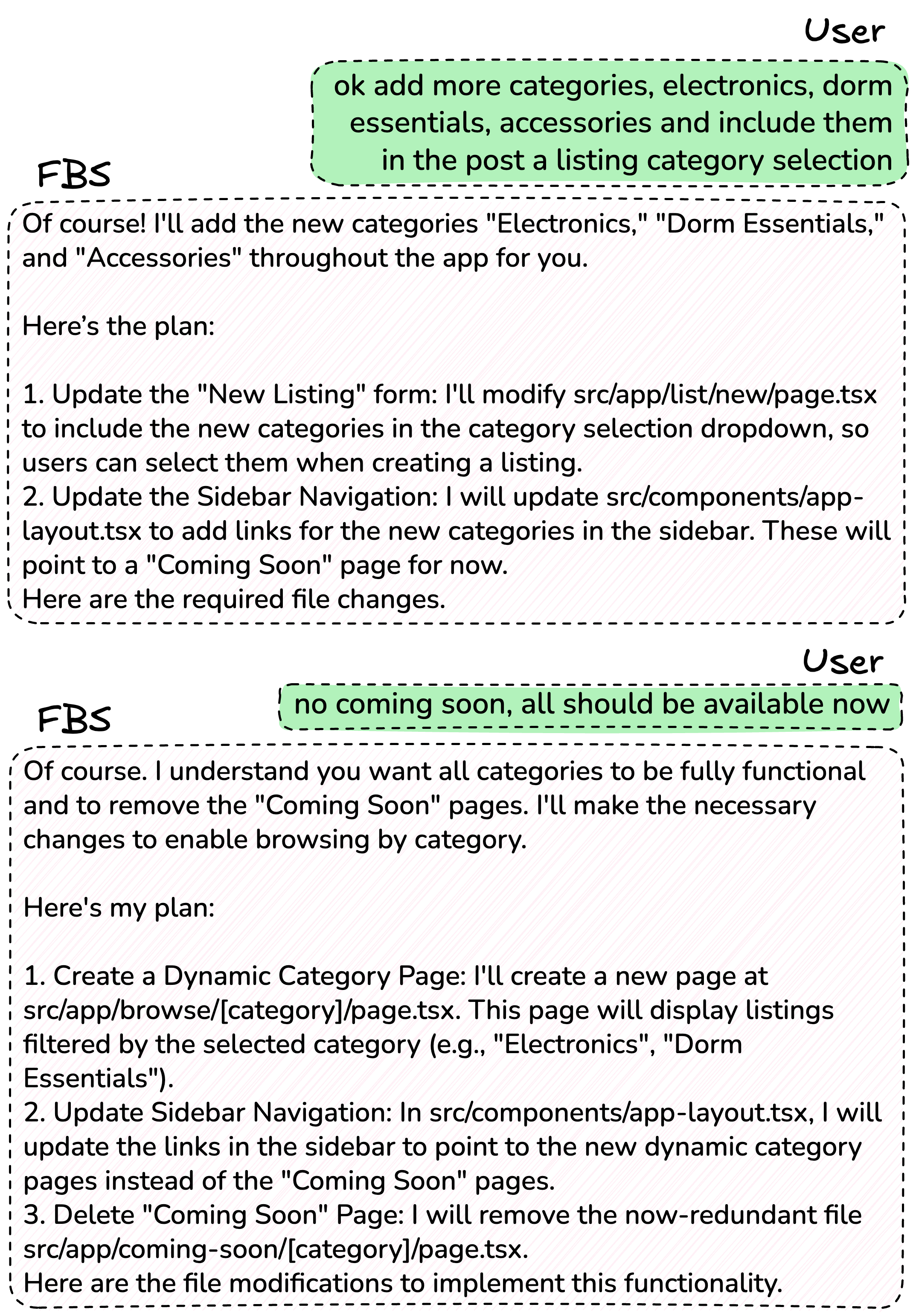}
    \caption{Iterative refinement prompt examples.}
    \label{fig:salma-squared-prompts}
\end{figure}

After that, the team used iterative refinement by prompting Firebase with the code itself. Many prompts after the initial generation included indications of where errors occurred or directly informed Firebase that there was a problem with a function, and Firebase worked to identify and remedy the mistake. Examples of the prompts can be seen in Figure \ref{fig:salma-squared-prompts}.

From an educational perspective, this case illustrates how vibe coding can support advanced learners working on complex, production-like systems. Rather than reducing the need for technical expertise, vibe coding redistributed effort toward architectural judgment, debugging, and design validation.

% \textcolor{red}{Things to discuss Jan 14:
% 1. What other kinds of cases we need? Following the 10 categories?
% 2. Missing GitHub repo / Access Issue
% 3. So hard reading other reports - all AI-generated long bullshits... Only read the prompt, but some were so hard to find. }

\begin{figure}[htbp]
    \centering
    \includegraphics[width=1\linewidth]{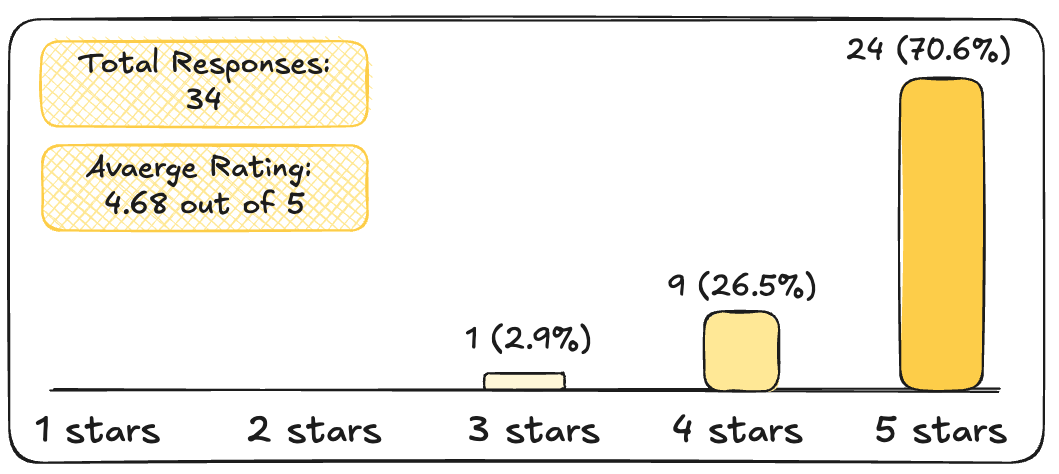}
    \caption{Distribution of ratings on overall helpfulness (1–5 Likert scale).}
    \label{fig:feedback_rating}
\end{figure}

\section{Reflection}

\subsection{The Participants' Post-hackathon Feedback}

\subsubsection{Overall Evaluation of the Hackathon}

Participants’ ratings of overall helpfulness were strongly positive (see Figure~\ref{fig:feedback_rating}). The mean rating was $M=4.68$ ($SD = 0.53$), with a median of 5, indicating both high satisfaction and low variability. A one-sample $t$-test comparing the mean rating to the neutral midpoint (3) confirmed that perceived helpfulness was significantly above neutral, $t(33) = 18.28$, $p < .001$, with a large effect size ($d = 3.13$). The 95\% confidence interval [$4.49$, $4.86$] further reflects strong consensus among participants. These results indicate that the hackathon was perceived as highly beneficial by the vast majority of participants, supporting its effectiveness as an educational mechanism.

\begin{table}[!t]
\centering
\caption{How the Competition Helped Participants}
\label{tab:helped_themes}
\renewcommand{\arraystretch}{1.15}
\begin{tabular}{
>{\centering\arraybackslash}p{2.2cm}
p{4cm}
c}
\toprule
\textbf{Theme} & \textbf{Summary} & \textbf{Count} \\
\midrule

LLM Prompting Skill Development &
Improved ability to structure prompts, iterate with LLM outputs, and control AI behavior during code generation. &
18 \\
\midrule

General Coding Skill Development &
Gains in programming knowledge, debugging skills, and coding confidence despite reliance on LLM assistance. &
14 \\
\midrule

Understanding LLM Capabilities and Limitations &
Increased awareness of what LLMs can and cannot do, emphasizing the need for human oversight. &
12 \\
\midrule

% Effective Use of AI and LLM Tools &
% Learning to use AI-assisted development tools and integrate them into workflows. &
% 10 \\
% \midrule

End-to-End Project Development &
Learning from building complete applications rather than isolated code snippets. &
7 \\
\midrule

Confidence, Enjoyment, and Motivation Boost &
Increased confidence and reduced intimidation toward coding and AI technologies. &
6 \\

\bottomrule
\end{tabular}

\vspace{0.3em}
\footnotesize
\textit{Note: Responses were coded using a multi-label qualitative approach; counts do not sum to the total number of responses.}
\end{table}

\begin{figure}[!t]
    \centering
    \includegraphics[width=0.75\linewidth]{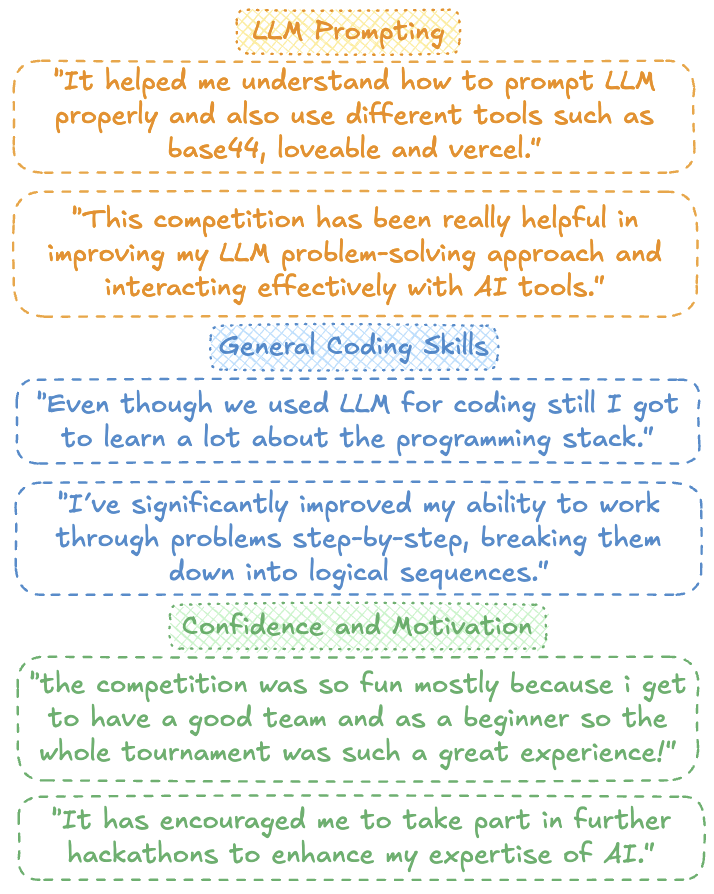}
    \caption{Representative participant comments grouped by thematic categories.}
    \label{fig:feedback_comments}
\end{figure}

We conducted a thematic analysis of open-ended responses using ChatGPT, followed by human review and minor refinements. The resulting thematic categories and frequency counts are summarized in Table \ref{tab:helped_themes}. Figure \ref{fig:feedback_comments} shows selected quotes of three categories from our collected feedback.

% First, prompt engineering and effective interaction with LLMs emerged as the dominant theme. Participants reported improved ability to structure prompts, iterate on outputs, and guide AI during code generation, indicating that the hackathon positioned prompt engineering as a substantive technical skill.

% Second, participants described meaningful gains in programming knowledge and system-level understanding, including debugging, integration, and deployment. Rather than diminishing foundational learning, AI-assisted coding appeared to shift effort toward evaluation, orchestration, and architectural reasoning.

% Third, many responses reflected a more nuanced understanding of LLM capabilities and limitations, emphasizing the need for human oversight and structured thinking. Improvements in problem decomposition and step-by-step reasoning were also frequently noted, suggesting that LLM-assisted development can scaffold higher-order reasoning.

% Finally, participants reported increased confidence, enjoyment, and motivation to pursue future AI-related projects. Overall, the findings indicate multidimensional learning, encompassing technical development, more calibrated understanding of AI systems, and positive shifts in mindset.

\subsubsection{Suggestions for Future Improvement}

Participants’ suggestions for future improvement provide further insight into how the competition could be refined (see Table \ref{tab:suggestion_themes}). 

\begin{table}[htbp]
\centering
\caption{Suggestions for Future Improvement}
\label{tab:suggestion_themes}
\renewcommand{\arraystretch}{1.2}
\begin{tabular}{
>{\centering\arraybackslash}p{2.2cm}
p{4 cm}
c}
\toprule
\textbf{Theme} & \textbf{Summary} & \textbf{Count} \\
\midrule

More Technical Guidance and Mentorship &
Requests for increased access to mentors, technical support, and structured guidance during development. &
12 \\
\midrule

Clearer Instructions and Evaluation Criteria &
Suggestions focused on improving clarity of rules, timelines, deliverables, and judging rubrics. &
10 \\
\midrule

Longer Duration or Extended Timeline &
Desire for more time to work on projects and iterate on ideas. &
9 \\
\midrule

More Workshops or Learning Sessions &
Requests for additional tutorials or sessions on LLM prompting, AI tools, and secure development practices. &
8 \\
\midrule

Improved Tooling and Infrastructure &
Suggestions to improve platforms, starter code, datasets, and technical reliability. &
7 \\
\midrule

More Feedback/ Post-Competition Review &
Desire for more detailed judge feedback and post-event evaluation opportunities. &
5 \\

\bottomrule
\end{tabular}

\vspace{0.5em}
\footnotesize
\textit{Note: Responses were coded using a multi-label qualitative approach; counts do not sum to the total number of responses.}
\end{table}

Overall, while satisfaction levels were high, participants’ constructive feedback highlights opportunities to enhance structure, rigor, and instructional depth without compromising the creativity and autonomy that characterized the event. These insights provide a foundation for refining future iterations of vibe coding–based educational hackathons. Incorporating structured mentorship and detailed post-competition feedback will be crucial to bridge the gap between rapid AI prototyping and rigorous software engineering practices.

% LIMITATION \& FUTURE WORK SECTION???? OR REFLECTION???? 

\section{Conclusion and Future Work}
Our research shows that vibe coding can be a way for students to break through coding education. Vibe coding does not require prior knowledge of the semantics of specific programming languages, enabling students to practice software engineering before officially taking a course and democratizing software creation. Redefining coding literacy becomes necessary to encompass more communication of logical steps and ideas than expertise in a specific programming language, especially in the age of LLMs. Furthermore, it exposes students to different aspects of the tech stack, such as version control systems, web hosting, and popular industry APIs. With students being able to define their own project and iterate on a product, coding becomes more fun and exciting. As AI-assisted coding becomes more commonplace in industry and education alike, earlier exposure to common AI tools becomes crucial for success in the software development field. 

However, vibe coding does not necessarily replace traditional computer science education. While the benefits of vibe coding allow students to rapidly ideate and test the potentials of AI and the product of what code can be, it does not replace studying different algorithms and best practices that make sure that applications run smoothly. When errors in the code happen, a person with a more traditional background will be able to smoothly identify the error and why it is an error, which is necessary for the longevity of a product. Also, vibe coding can lead to false confidence as vibe coders might be inclined to trust an LLM despite security and functional errors in the code generated. A traditional software developer would be required to help debug and secure the code. As of now, LLMs are still not developed enough to exist autonomously from traditional software developers.

To extend this work, a couple of considerations can be made.
\begin{enumerate}
    \item \textbf{More vibe coding hackathons}. This study represents a small-scale hackathon. The hackathon also spanned only one month whereas more asynchronous hackathons can be longer-term. For example, a year-long hackathon can be conducted to receive more complex projects.
    \item \textbf{Expert group}. Our hackathon mainly received university students who have been studying for 1-2 years. A couple participants were freshly graduated from university. We did not have a participant who belongs to an expert group, for example, a senior AI engineer or a software engineer with 20+ years of experience.
    \item \textbf{Different areas of education}. Vibe coding represents only one sector for AI inside of education. More exploration can be done in terms of different subject matters such as mathematics and social science.
\end{enumerate}
% - more hackathons + longer hackathon (more complex, industrial level) + year-long

% - most competitors are using human-in-the-loop and look into how experts building can be one of the future directions

% - vibe exam

\bibliographystyle{IEEEtran}      
\bibliography{ref}

\end{document}